\documentclass[lettersize,journal]{IEEEtran}
\usepackage{amsmath,amsfonts}
\usepackage{algorithmic}
\usepackage{array}
\usepackage[caption=false,font=normalsize,labelfont=sf,textfont=sf]{subfig}
\usepackage{textcomp}
\usepackage{stfloats}
\usepackage{url}
\usepackage{verbatim}
\usepackage{graphicx}
\def\BibTeX{{\rm B\kern-.05em{\sc i\kern-.025em b}\kern-.08em
    T\kern-.1667em\lower.7ex\hbox{E}\kern-.125emX}}
\usepackage{balance}

\newtheorem{theorem}{Theorem}
\newtheorem{corollary}{Corollary}
\newtheorem{proposition}{Proposition}

\usepackage{makecell}
\usepackage{threeparttable}
\usepackage{booktabs}
\usepackage{amssymb}
\usepackage{cuted}
\usepackage{multirow}
\usepackage{float}
\usepackage{placeins}

\begin{document}
\title{Towards a Unified Theoretical Framework \\ for Splitting-based Self-Supervised MRI Reconstruction}
\author{Siying Xu, Kerstin Hammernik, Daniel Rueckert \IEEEmembership{(Fellow, IEEE)},\\ Sergios Gatidis, and Thomas Küstner \IEEEmembership{(Member, IEEE)}
\thanks{This work was supported by the Deutsche Forschungsgemeinschaft (DFG, German Research Foundation) under Germany’s Excellence Strategy – EXC 2064/1 – Project number 390727645.}
\thanks{Siying Xu, Sergios Gatidis and Thomas Küstner are with Medical Image and Data Analysis (MIDAS.lab), Department of Diagnostic and Interventional Radiology, University of Tuebingen, Tuebingen, Germany (e-mail: \{siying.xu; marcel.frueh; sergios.gatidis; thomas.kuestner\}@med.uni-tuebingen.de). }
\thanks{Kerstin Hammernik and Daniel Rueckert are with Chair for AI in Healthcare and Medicine, Technical University of Munich (TUM) and TUM University Hospital, Munich, Germany (e-mail: \{k.hammernik; daniel.rueckert\}@tum.de). }
\thanks{Daniel Rueckert is also with the Department of Computing, Imperial College London, London, United Kingdom.}
\thanks{Sergios Gatidis is also with the Department of Radiology, Stanford University, Stanford, California, USA}}

\markboth{Journal of \LaTeX\ Class Files,~Vol.~XX, No.~XX, XX~2025}%
{Towards a Unified Theoretical Framework for Self-Supervised MRI Reconstruction}

\maketitle

\begin{abstract}
The demand for high-resolution, non-invasive imaging continues to drive innovation in magnetic resonance imaging (MRI), but long acquisition times remain a major practical limitation. Although deep learning-based reconstruction methods have enabled accelerated imaging, their predominant supervised paradigm relies on fully-sampled reference data that are difficult to acquire in practice. Self-supervised learning (SSL) has therefore emerged as a promising alternative, among which splitting methods are a widely used strategy. However, most existing splitting-based methods are empirically designed, and a unified theoretical understanding remains limited. In this work, we introduce UNITS (\textit{\textbf{Uni}}fied \textit{\textbf{T}}heory for \textit{\textbf{S}}plitting-based self-supervision), a general theoretical framework for splitting-based self-supervised MRI reconstruction. Theoretically, we show that the self-supervised risk can be expressed as a weighted supervised risk. Consequently, self-supervision admits the same pointwise Bayes-optimal predictor as supervised learning. We further relate the training residual to the prediction bias, revealing how different sampling mechanisms affect training behavior. UNITS makes a broad class of existing methods interpretable as special cases within a common framework, and provides a general design space through sampling stochasticity and flexible data utilization. Together, these contributions establish UNITS as a theoretical foundation, a practical paradigm, and a benchmark for interpretable, generalizable, and applicable self-supervised MRI reconstruction. 
\end{abstract}


\section{Introduction}
\IEEEPARstart{M}{edical} imaging is an indispensable, non-invasive tool in clinical diagnostics. Among its various modalities, magnetic resonance imaging (MRI) has long been a cornerstone owing to its excellent soft-tissue contrast and absence of ionizing radiation. However, the inherently long acquisition time of MRI poses critical limitations, including patient discomfort, increased sensitivity to motion artifacts, and reduced scanning throughput. To accelerate MRI acquisition, a widely adopted strategy is to undersample the k-space data, the acquisition domain of MRI, and reconstruct the image by exploiting prior knowledge such as coil sensitivities and transform-domain sparsity. Among the most widely used approaches are parallel imaging (PI)~\cite{pruessmann1999sense, griswold2002generalized, lustig2010spirit, uecker2014espirit} and compressed sensing (CS)~\cite{donoho2006compressed, lustig2007sparse, lustig2008compressed, ye2019compressed}. PI exploits the spatial sensitivity profiles of multiple receiver coils to acquire data in parallel, while CS exploits sparse representations combined with randomized sampling and non-linear reconstruction. Despite their effectiveness, these traditional methods typically involve iterative algorithms and handcrafted priors, limiting the achievable acceleration rates.

Recently, deep learning (DL) has started to revolutionize MRI reconstruction by leveraging data-driven priors to improve both reconstruction efficiency and image quality~\cite {wang2016accelerating, hammernik2018learning, kustner2020cinenet, hammernik2023physics, heckel2024deep, xu2025attention}. Most existing approaches follow a \textit{supervised learning} paradigm, training reconstruction networks on pairs of undersampled data and fully-sampled images. This strategy requires large-scale fully-sampled datasets, which are challenging to acquire in practice, particularly in dynamic imaging, where prolonged scans are highly susceptible to motion artifacts caused by breathing or other involuntary movements. Public datasets like fastMRI~\cite{zbontar2018fastmri}, OCMR~\cite{chen2020ocmr}, and CMRxRecon~\cite{wang2024cmrxrecon} provide valuable resources, but remain limited in anatomical diversity and contrast settings. Furthermore, many datasets considered "fully-sampled" in clinical settings are in fact mildly accelerated and reconstructed using traditional methods such as PI or CS, causing the nominal ground truth to inherit algorithmic biases and artifacts~\cite{sartoretti2018common}. This reliance on imperfect references constrains the attainable performance of supervised models. As such, it is of increasing interest for learning paradigms to avoid the dependence on fully-sampled data.

\textit{Self-supervised learning (SSL)} approaches~\cite{yaman2020self, yaman2021zero, cui2022self, yaman2022multi, zhou2022dual, cho2023improved, korkmaz2023self, molaei2023implicit, wang2023kband, huang2025subspace, xu2025self, yu2025bilevel, li2025self} have recently gained traction as a promising solution to the scarcity of fully-sampled data. Existing methods can be broadly grouped into four categories: (i) Splitting-based methods~\cite{yaman2020self, yaman2022multi, zhou2022dual, wang2023kband} partition the acquired undersampled k-space into subsets, used for network input and for the training loss. (ii) Subject-specific or zero-shot learning~\cite{yaman2021zero, cho2023improved}, wherein a single scan is further split into training input, loss, and validation subsets for per-scan tuning without external datasets. (iii) Implicit neural representations (INR)~\cite{molaei2023implicit, huang2025subspace, yu2025bilevel}, which learn in a zero-shot or few-shot fashion coordinate-based mapping from undersampled data. (iv) Generative approaches~\cite{korkmaz2023self, cui2022self} learn priors directly from the undersampled data using generative models.

Among these categories, splitting-based methods are particularly common because they provide a simple and practical way to construct both network inputs and supervision signals directly from the undersampled acquisitions. Despite their empirical success, existing methods are often introduced with different choices of sampling masks, subset assignments, and loss construction. As a result, even methods that follow a similar reconstruction principle remain difficult to interpret and organize within a common conceptual framework, and their design choices are not systematically understood.

Recent studies have addressed these issues in complementary ways. SSIBench~\cite{wang2025benchmarking} enables standardized and reproducible comparisons of self-supervised feedforward methods by evaluating loss functions under a controlled experimental setup in which the other pipeline components are fixed. Theoretical studies have analytically clarified specific splitting-based formulations, including SSDU through a Noisier2Noise-based analysis~\cite{millard2023theoretical} and k-band through a stochastic gradient descent perspective~\cite{wang2023kband}. Nevertheless, these efforts remain partial in scope: SSIBench is aimed at systematic empirical comparison rather than conceptual and workflow unification, and current theoretical studies are tied to particular method formulations. A framework that both organizes splitting-based methods under a common formalism and provides a principled theoretical interpretation is still missing.

In this work, we propose UNITS (\textbf{\textit{uni}}fied \textbf{\textit{t}}heory for \textbf{\textit{s}}plitting-based self-supervision), a general framework for splitting-based self-supervised MRI reconstruction methods. UNITS provides a common formalism for describing different splitting-based methods within a single conceptual framework, despite their differences in sampling mechanisms, network architectures, and loss functions. Within this framework, \textit{sampling stochasticity} and \textit{flexible data utilization} serve as two key design dimensions that capture methodological variation across existing approaches and support the development of new variants.

More importantly, UNITS establishes a risk-based characterization of splitting-based self-supervision. We show that, under arbitrary pointwise losses, the self-supervised risk can be expressed as a weighted supervised risk, with the weights induced by the re-undersampling mechanism. This characterization implies that splitting-based self-supervision admits the same pointwise Bayes-optimal predictor as in supervised learning under the same loss. Thereby, UNITS provides a principled explanation for the empirical success of this family of methods. We further analyze the deviation of the learned estimator from this Bayes-optimal solution, offering additional theoretical insight, beyond previous works~\cite{millard2023theoretical, moran2020noisier2noise}, into how different sampling mechanisms influence training behavior. Taken together, these contributions position UNITS as a unified \textit{conceptual} and \textit{theoretical} foundation for understanding splitting-based SSL methods and guide the design of future ones. 

\section{Methods}
\noindent UNITS provides a general framework for splitting-based self-supervised MRI reconstruction, which is agnostic to acquisition sequences, sampling patterns, and network architectures, aiming to unify learning strategies at a conceptual level. The overall workflow is illustrated in Fig.~\ref{fig:Framework}.

\subsection{Workflow}
The framework consists of three stages: (a) initial undersampling, which can be performed prospectively (i.e., acquisition of undersampled data) or retrospectively (i.e., undersampling of fully-sampled data) to enable broad applicability across diverse sampling scenarios; (b) self-supervised training via re-undersampling, where multiple masks are applied to generate subsets of the acquired k-space that serve as inputs or supervision (loss); and (c) inference, where the trained network directly reconstructs images from undersampled acquisitions.

During training, the initially acquired k-space is further re-undersampled by applying multiple masks $M_{1},\dots,M_{L}(L\ge 2)$ at each step, generating multiple subsets $y_{1},\dots,y_{L}$, each containing a different portion of the acquired data. These subsets can be flexibly assigned as network inputs or supervision signals, with the requirement that loss is always computed between different subsets. Input subsets are passed through the reconstruction network, which can operate directly in k-space or in the image domain after applying the adjoint forward operator. In loss calculation, the reconstructed k-space is compared with the sampled entries of the supervision subsets. In this way, the network is optimized without requiring any fully-sampled data.

\begin{figure*}[t]
\centering
\includegraphics[width=\linewidth]{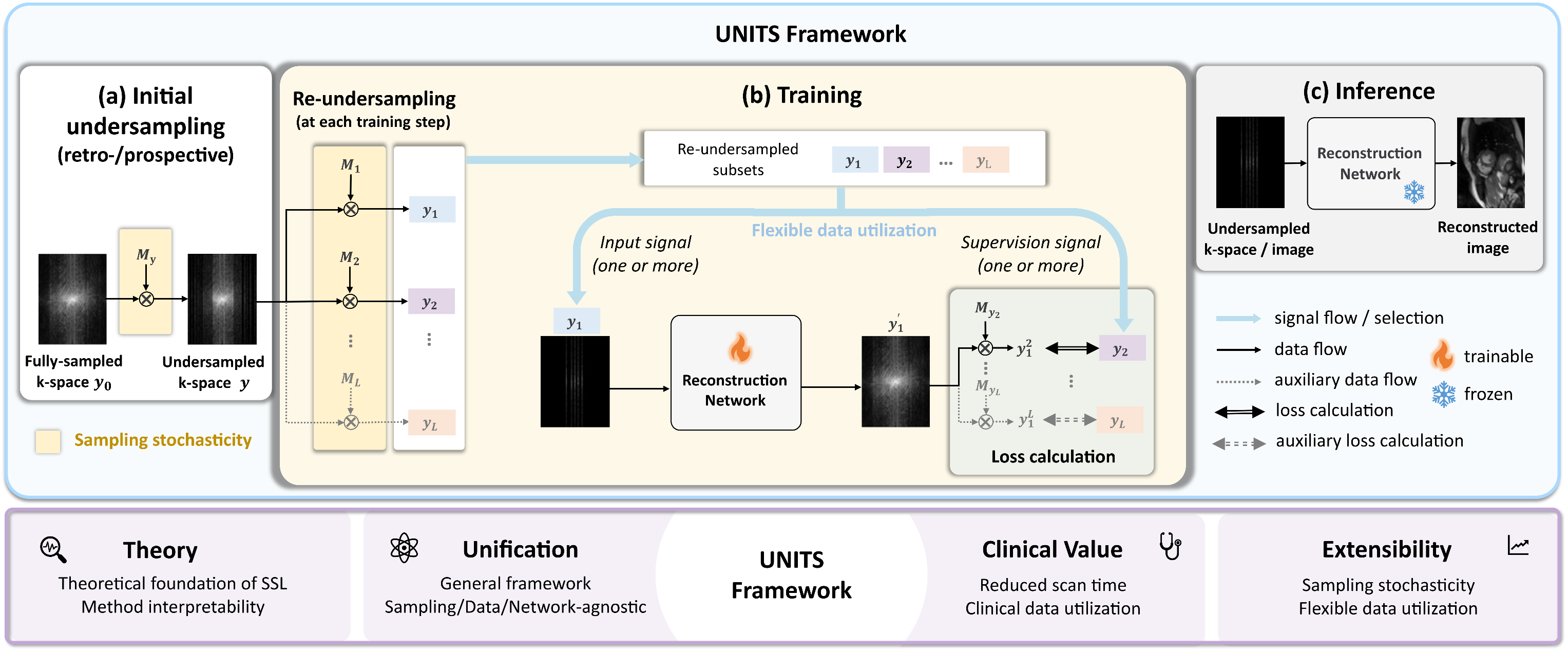}
\caption{\textbf{Overview of the proposed UNITS (unified theory for splitting-based self-supervision) framework.} The framework defines a general splitting-based self-supervised learning paradigm for MRI reconstruction: (a) Initial undersampling: an undersampling mask $M_{y}$ is applied to acquire k-space $y$ for training. (b) Training: at each step, $y$ undergoes re-undersampling with multiple random masks $M_{1},\dots,M_{L}(L\ge 2)$ to generate subsets $y_{1},\dots,y_{L}$, which are flexibly assigned as inputs or supervision signals. Input subsets are processed by the reconstruction network and compared in loss calculation with measured entries in the supervision subsets that differ from the input. Solid arrows denote the main training pathways that are mandatory for learning, while dashed arrows indicate auxiliary pathways that support framework generality and extensibility. (c) Inference: the trained network directly reconstructs images from undersampled data. The bottom panel summarizes the core advantages of the UNITS framework.}
\label{fig:Framework}
\end{figure*}

Two core design elements make UNITS a generalizable framework that subsumes diverse splitting-based SSL strategies as special cases. First, sampling stochasticity (Section~\ref{sec:method_stochasticity}) allows arbitrary sampling patterns in both the initial and re-undersampling stages. Auxiliary pathways (dashed arrows in Fig.~\ref{fig:Framework}) further support the construction of multiple subsets with distinct sampling characteristics, enriching both input and supervision signals. Second, flexible data utilization (Section~\ref{sec:units_cross}) permits subsets to be assigned across inputs and losses, allowing the network to process multiple inputs in parallel and accommodate multiple loss terms, thereby maximizing the use of available sampling information without modifying the network architecture. To demonstrate these principles, we instantiate the framework in two variants: \textit{UNITS-Base} (Section~\ref{sec:method_stochasticity}), which incorporates sampling stochasticity, and \textit{UNITS-Cross} (Section~\ref{sec:units_cross}), which extends it with flexible data utilization.

\subsection{Risk Characterization of Splitting-based Self-supervision}
\label{sec:risk_characterization}
This section characterizes the population risk of splitting-based self-supervision under the UNITS framework. The theoretical analysis focuses on the main pathways (solid arrows in Fig.~\ref{fig:Framework}). Without loss of generality, the same formulation applies to the optional auxiliary pathways (dashed arrows) by analogous derivation. 

We formalize splitting-based self-supervised reconstruction within the UNITS framework. Mathematically, we use uppercase letters (e.g., $Y$, $Y_{0}$) to denote random variables in the theoretical formulation, while their lowercase counterparts (e.g., $y$, $y_{0}$) represent specific realizations as used in experiments and figures. An overview of the used variables is presented in Appendix Table~\ref{tab:notation}.

Let $Y_{0}\in \mathbb{C}^{N}$ denote the unknown fully-sampled k-space, stacked into a one-dimensional vector of length $N$, where $N$ represents the total number of k-space samples across all dimensions (e.g., $N=N_xN_yN_zN_tN_c$ for 3D spatial, temporal, and coil dimensions). This formulation is dimension-agnostic and applies without loss of generality to arbitrary acquisitions. We observe an undersampled k-space $Y\in \mathbb{C}^{N}$, which is acquired from $Y_{0}$ with the initial undersampling mask $M_{Y}$:
\begin{equation}
    Y=M_{Y}\odot Y_{0}.
\end{equation}
Here, $\odot$ denotes the Hadamard (element-wise) product. $M_{Y}\in\left \{ 0,1 \right \}^{N}$ is a point-wise binary mask, where the probability of a specific location $i$ being sampled $(M_{Y, i}=1)$ is $p_{i}\in (0,1]$. 

During training, the undersampled k-space $Y$ serves as the starting point. To enable self-supervised learning, we re-undersample $Y$ by two random masks, denoted by the random variables $M_{1},M_{2}\in\left \{ 0,1 \right \}^{N}$:
\begin{equation}
    \begin{split}
        Y_{1}=M_{1}\odot Y=(M_{1}\odot M_{Y})\odot Y_{0}=M_{Y_{1}}\odot Y_{0},\\
        Y_{2}=M_{2}\odot Y=(M_{2}\odot M_{Y})\odot Y_{0}=M_{Y_{2}}\odot Y_{0},
    \end{split}
    \label{eq:re_undersampling}
\end{equation}
where the probability of a sampled point $Y_i$ being re-sampled by $M_{1}$ and $M_{2}$ are denoted as $q_{i}$ and $r_{i}$, respectively, i.e., $q_{i}=P[Y_{1,i}\ne 0\mid Y_{i}\ne 0]$, $r_{i}=P[Y_{2,i}\ne 0\mid Y_{i}\ne 0]$. We define the effective undersampling masks as:
\begin{equation}
    \begin{split}
        M_{Y_{1}}=M_{1}\odot M_{Y},\\
        M_{Y_{2}}=M_{2}\odot M_{Y}.
    \end{split}
    \label{eq:effective_mask}
\end{equation}
As a result, the re-undersampling process produces two further undersampled k-spaces $Y_{1}\in \mathbb{C}^{N}$ and $Y_{2}\in \mathbb{C}^{N}$.

Let $f:\mathbb{C}^{N}\to\mathbb{C}^{N}$ denote a reconstruction network. During training, one re-undersampled realization (e.g., $Y_{1}$) is used to construct the network input, and another re-undersampled realization (e.g., $Y_{2}$) is used for self-supervision. The network is therefore trained to minimize the self-supervised population risk:
\begin{equation}
    \mathcal{R}_{\mathrm{ssl}}(f)=\mathbb{E}\left[L(M_{Y_2}\odot f(Y_1),\,Y_2)\right],
    \label{eq:risk_ssl}
\end{equation}
where:
\begin{equation}
    L(A,B)=\sum_{i=1}^{N}\ell (A_{i},B_{i}),
    \label{eq:pointwise_loss_func}
\end{equation}
and $\ell:\mathbb{C}\times \mathbb{C}\to\mathbb{R}_{+}$ is a pointwise loss functions, such as the $\ell_{1}$ or $\ell_{2}$ loss. Here, $\mathbb{E}[\cdot]$ denotes the expectation over all random variables within the bracket. 

For comparison, the corresponding supervised risk under the same effective input $Y_{1}$ and the same loss function is:
\begin{equation}
    \mathcal{R}_{\mathrm{sup}}(f)=\mathbb{E}\left[L(f(Y_1),\,Y_0)\right].
    \label{eq:risk_sup}
\end{equation}

\vspace{0.5em}
\begin{theorem}[\textbf{Self-supervised Risk as a Weighted Conditional Supervised Risk}]\label{thm:equivalence}
Under a pointwise loss, the self-supervised risk in Eq.~\eqref{eq:risk_ssl} can be written as a weighted supervised risk:
\begin{equation}
    \mathcal{R}_{\mathrm{ssl}}(f)=\mathbb{E}\left[\sum_{i=1}^{N}w_{i}(Y_{1}) \mathbb{E}\left[\ell (f_{i}(Y_1),Y_{0,i})\mid Y_1\right]\right],
\end{equation}
where $w_i(Y_1):=\mathbb{E}[M_{{Y_2},i}\mid Y_1]$ denotes the conditional supervision weight induced by the re-undersampling mechanism. The proof of Theorem~\ref{thm:equivalence} is provided in Appendix~A.
\end{theorem}
\vspace{0.5em}

Theorem~\ref{thm:equivalence} shows that splitting-based self-supervised MRI reconstruction can be interpreted as minimizing a weighted conditional supervised risk, where the weights $w_{i}(Y_1)$ quantify how likely each k-space location is to be available for supervision under the chosen splitting strategy. Since $w_{i}(Y_1)$ is determined solely by the sampling mechanism, it admits an explicit form under the UNITS framework.

\vspace{0.5em}
\begin{proposition}[\textbf{Explicit Form of the Conditional Supervision Term}]\label{proposition:wi(Y1)}
    Under the UNITS framework, the conditional supervision weight $w_{i}(Y_1)$ can be explicitly written as:
    \begin{equation}
    \resizebox{\linewidth}{!}{$
            w_{i}(Y_{1})=(1-\frac{1-p_{i}}{1-p_{i}q_{i}}(1-\mathbb{E}[M_{Y,i}M_{1,i}\mid Y_{1}])) \mathbb{E}[M_{2,i}\mid Y_{1}, Y_{i}\ne 0],
    $}
        \label{eq:wi}
    \end{equation}
    The detailed derivation is provided in Appendix~B.
\end{proposition}
\vspace{0.5em}

Proposition~\ref{proposition:wi(Y1)} gives the explicit form of $w_{i}(Y_1)$. Under the positive condition $w_{i}(Y_1)>0$ (i.e., $p_{i}>0$, $q_i<1$, and $r_i>0$), this sampling-dependent weighting preserves the set of pointwise minimizers, as stated below.

\vspace{0.5em}
\begin{proposition}[\textbf{Pointwise Equivalence of Bayes-optimal Predictor}]\label{proposition:equivalence}
    If $w_{i}(Y_1)>0$ for all $i$, then multiplying each pointwise conditional risk term by $w_{i}(Y_1)$ does not change its set of pointwise minimizers. Consequently, the weighted self-supervised risk induces the same Bayes-optimal predictor as the corresponding supervised loss.
\end{proposition}
\vspace{0.5em}

Importantly, this equivalence is established at the population and Bayesian levels, rather than implying full equivalence in the parameter optimization of the neural networks. Proposition~\ref{proposition:equivalence} reveals that for each k-space location, self-supervision points to a valid supervised Bayes predictor, which identifies the correctness of the UNITS framework. Based on this proposition, the resulting Bayes-optimal predictors under commonly used losses can be characterized as follows:

\vspace{0.5em}
\begin{corollary}[\textbf{$\ell_1$ loss}]\label{corollary:mae}
    If $\ell(a,b)=\left | a-b \right |$ and $w_{i}(Y_1)>0$ for all $i$, then the pointwise Bayes-optimal predictor of UNITS satisfies:
    \begin{equation}
        f_{i}^{\star}(Y_1)\in \arg\min_{a}\,\mathbb{E}\bigl[\lvert a - Y_{0,i}\rvert \mid Y_1\bigr].
    \end{equation}
\end{corollary}
\vspace{0.5em}

\vspace{0.5em}
\begin{corollary}[\textbf{$\ell_2$ loss}]\label{corollary:mse}
    If $\ell(a,b)=\left | a-b \right |^{2}$ and $w_{i}(Y_1)>0$ for all $i$, then the pointwise Bayes-optimal estimator of UNITS is:
    \begin{equation}
        f_{i}^{\star}(Y_1)=\mathbb{E}[Y_{0,i}\mid Y_1].
        \label{eq:optimal_predictor_l2}
    \end{equation}
    Equivalently, stacking all coordinates yields the vector-form estimator:
    \begin{equation}
        f^{\star}(Y_1)=\mathbb{E}[Y_{0}\mid Y_1].
    \end{equation}
\end{corollary}
\vspace{0.5em}

In summary, Theorem~\ref{thm:equivalence} and Proposition~\ref{proposition:equivalence} provide the general theoretical interpretation of UNITS, whereas Corollaries~\ref{corollary:mae} and~\ref{corollary:mse} show how this interpretation specializes to commonly used losses. In particular, the Bayes-optimal predictor under $\ell_2$ admits a closed-form characterization as the conditional mean, which serves as the starting point for the bias analysis in the following section.

\subsection{Bias Analysis}
\label{sec:bias_analysis}
Corollaries~\ref{corollary:mae} and~\ref{corollary:mse} characterize the Bayes optimal predictor under ideal conditions. In practice, however, limited data availability, finite model capacity, and imperfect optimization lead to a non-zero residual between the learned predictor and the optimal solution. To explicitly characterize how re-undersampling affects training behavior in this non-ideal regime, we next analyze the training residual under the UNITS framework. 

Given the network input $Y_1$, the point-wise training residual at k-space location $i$ is:
\begin{equation}
    \mathbb{E}[\mathcal{E}_{i}\mid Y_{1}]=\mathbb{E}[(M_{Y_{2},i}\cdot f_i(Y_{1})-Y_{2,i}) \mid Y_{1}].
    \label{eq:training_residual}
\end{equation}
While this residual is the quantity directly exposed during training, the more relevant measure of estimator quality is the deviation of the learned predictor from the Bayes-optimal predictor, which we defined as the prediction bias:
\begin{equation}
    e_{i}(Y_1):=f_i(Y_1)-f^{\star}_i(Y_1).
    \label{eq:def_ei}
\end{equation}
The following proposition establishes the relation between the training residual and the prediction bias.

\vspace{0.5em}
\begin{proposition}[\textbf{Factorization of Training Residual}]\label{proposition:feactorization}
    Under the UNITS framework with $\ell_2$ loss, the conditional expected training residual at k-space location $i$ satisfies:
    \begin{equation}
        \mathbb{E}[\mathcal{E}_{i}\mid Y_1]=w_{i}(Y_1)\cdot e_{i}(Y_1),
        \label{eq:factorization}
    \end{equation}
    where
    \begin{equation}
        e_{i}(Y_1)=f_i(Y_1)-\mathbb{E}[Y_{0,i}\mid Y_{1}],
    \end{equation}
    and
    \begin{equation}
        w_i(Y_1)=\mathbb{E}[M_{{Y_2},i}\mid Y_1].
    \end{equation}
    The proof is given in Appendix~C.
\end{proposition}
\vspace{0.5em}

This factorized formulation shows that, under $\ell_2$ loss, the training residual is a weighted combination of the prediction bias. The weights $w_{i}(Y_1)$ in Eq.~\ref{eq:wi} are determined by the sampling process and govern how prediction bias propagates into the training error. Therefore, different sampling mechanisms influence the training dynamics through distinct behaviors of $w_{i}(Y_1)$.

To compare different re-undersampling mechanisms at the level of expected training behavior, we further average Eq.~\eqref{eq:factorization} over the randomness of the input subset. Here, we examine two common re-undersampling strategies: the disjoint re-undersampling~\cite{yaman2020self, yaman2021zero, yaman2022multi, zhou2022dual, cho2023improved, korkmaz2023self}, and the independent re-undersampling~\cite{xu2025self}.

\vspace{0.5em}
\begin{corollary}[Expected training residual under disjoint re-undersampling]\label{corollary:disjoint}
    Under disjoint re-undersampling, each acquired k-space entry in $Y$ is assigned exclusively to either the input subset $Y_{1}$ or the supervision subset $Y_{2}$, i.e., $M_1\cap M_2=\emptyset$ and $M_1\cup M_2=M_Y$. The expected training residual in this case is:
    \begin{equation}
        \mathbb{E}[\mathcal{E}_{i}]=p_i(1-q_i)\cdot\mathbb{E}[e_{i}(Y_{1})\mid Y_{1,i}=0].
    \end{equation}
\end{corollary}
\vspace{0.5em}

\vspace{0.5em}
\begin{corollary}[Expected training residual under independent re-undersampling]\label{corollary:independent}
    Under independent re-undersampling, the re-undersampling masks $M_{1}$ and $M_{2}$ are sampled independently given the initial acquisition $Y$. In this case, the expected training residual is:
    \begin{equation}
    \resizebox{\linewidth}{!}{$
        \mathbb{E}[\mathcal{E}_{i}]=p_ir_i(1-q_i)\mathbb{E}[e_{i}(Y_{1})\mid Y_{1,i}=0]+p_iq_ir_i\mathbb{E}[e_{i}(Y_{1})\mid Y_{1,i}\ne0].
    $}
    \end{equation}
    Derivations of Corollaries~\ref{corollary:disjoint} and~\ref{corollary:independent} are provided in Appendix D.
\end{corollary}
\vspace{0.5em}

Corollaries~\ref{corollary:disjoint} and~\ref{corollary:independent} show the expected residual signal seen during training under two different re-undersampling mechanisms. In the disjoint case, only locations excluded from the input subset contribute to the expected residual. By contrast, independent re-undersampling preserves an additional residual pathway on locations already included in the input subset. Consequently, disjoint splitting suppresses supervision on frequent input-visible locations, whereas independent splitting retains supervision contributions from both input-hidden and input-visible locations. This distinction provides a theoretical basis for the stochastic sampling choices introduced in the following section.

\subsection{Sampling Stochasticity}
\label{sec:method_stochasticity}
The risk characterization in Section~\ref{sec:risk_characterization} shows that the sampling mechanism determines the effective supervision weights $w_i(Y_1)$ and resulting learning objective. The bias analysis further shows how the sampling probabilities control the contribution of the prediction bias to the training residual. Therefore, sampling is not merely a data-preparation step, but also an important design component that influences training dynamics. To support flexible sampling strategies, UNITS introduces \textit{sampling stochasticity} as a key component for describing, comparing, and extending diverse sampling mechanisms in splitting-based SSL methods. Here, we identify three specific sampling degrees of freedom that UNITS can accommodate:

\subsubsection{Initial undersampling randomness} 
The UNITS framework supports both retrospective and prospective initial undersampling. In retrospective settings, the initially undersampled k-space $Y$ is generated by applying an undersampling mask $M_{Y}$ to the fully-sampled acquisition. In this scenario, UNITS allows $M_{Y}$ to vary across training steps. Specifically, $M_{Y}$ can be drawn from a prescribed distribution (e.g., Gaussian or Bernoulli) with \textit{random acceleration rates} (e.g., between $R=2$ and $R=16$) and \textit{random generation seeds}. This formulation can naturally extend to prospective studies, where data acquired under different acceleration rates could be jointly used for network training. Such flexibility has the potential to relax dataset constraints and enhance the utility of numerous clinical undersampled datasets.

\subsubsection{Re-undersampling ratio variability} Rather than fixing the proportion of points selected in the re-undersampling masks, UNITS allows \textit{random re-undersampling ratios} at each training step. This variation induces changes in the effective acceleration rate of re-undersampled subsets, increasing the diversity of training inputs and supervision signals. Similar to dropout or data augmentation, such stochasticity acts as a form of implicit regularization, helping the network to avoid overfitting to fixed sampling patterns and to improve performance under distribution shifts.

\subsubsection{Independent subsets sampling} UNITS does not require the re-undersampling masks to be disjoint or deterministically related. For any location in the initially acquired k-space $Y$, its inclusion in $Y_{1}, Y_{2},\dots, Y_{L}$ can be determined independently. As a result, input and supervision locations vary across training iterations, which further expands the diversity of input and supervision signals and encourages the network to generalize beyond fixed loss regions.

To demonstrate how these sampling stochasticities can be jointly utilized in practice, we implement a baseline model within the UNITS framework, termed \textit{UNITS-Base} (Supplementary Fig.~1). At each training step, \textit{UNITS-Base} randomly selects an initial acceleration rate and independently draws re-undersampled subsets from the acquired k-space with varying locations and sampling ratios. For simplicity, we generate only two subsets to demonstrate the practical feasibility of the framework: one for constructing the input and the other for the supervision signal.

Importantly, as long as each location has a positive probability of being selected for supervision, the resulting training objective preserves the same pointwise Bayes-optimal predictor under the risk characterization in Section~\ref{sec:risk_characterization}. In practice, sampling stochasticity makes it possible to incorporate different mask distributions, acceleration rates, re-undersampling ratios, and subset-dependence structures within a common framework. Together with the flexible data utilization introduced in the next section, sampling stochasticity forms the basis of UNITS as a generalizable reconstruction pipeline.

\subsection{Flexible Data Utilization}
\label{sec:units_cross}
Sampling stochasticity controls how re-undersampled subsets are generated, while flexible data utilization controls how these subsets are assigned to network inputs and supervision signals. The UNITS framework accommodates \textit{multiple} inputs and loss terms, enabling complementary supervision between different sampling realizations, thereby maximizing the utilization of available sampling information without modifying the reconstruction architecture. 

To demonstrate this flexibility, we extend \textit{UNITS-Base} with a cross-consistency loss, resulting in an instantiation termed \textit{UNITS-Cross} (Supplementary Fig.~2). In \textit{UNITS-Cross}, the network is trained to predict $Y_2$ from $Y_1$ and, simultaneously, to predict $Y_1$ from $Y_2$ using the same shared parameters. Formally, the cross-consistency risk is defined as:
\begin{equation}
\resizebox{\linewidth}{!}{$
    \mathcal{R}_{\mathrm{cross}}(f)=\frac{1}{2}\mathbb{E}\left[L(M_{Y_2}\odot f(Y_1),\,Y_2)\right]+ \frac{1}{2}\mathbb{E}\left[L(M_{Y_1}\odot f(Y_2),\,Y_1)\right],
$}
\label{eq:cross_kspace_loss}
\end{equation}
where $L$ is the pointwise loss in Eq.~\eqref{eq:pointwise_loss_func}, and $f(\cdot)$ denotes the shared reconstruction network.

Although the UNITS framework in principle allows generating more than two re-undersampled subsets, in \textit{UNITS-Cross}, we restrict this number to two as a practical trade-off between data utilization and computational cost. Each additional subset would require an extra forward and backward pass through the network, substantially increasing computational cost. 

Importantly, the same conceptual and theoretical formalisms developed in Section~\ref{sec:risk_characterization} hold for the auxiliary pathway. In addition to preserving the convergence guarantee, the cross-consistency loss also provides a statistical motivation. By averaging supervision across two re-undersampled pathways, it may reduce the variance of the prediction bias, thereby yielding a potential stabilizing benefit in practice. Appendix~E analyzes this effect through variance decomposition of the averaged prediction bias. The analysis shows that the potential benefit depends on the correlation between the two pathways: weakly correlated errors can lead to variance reduction, whereas strongly correlated errors reduce this benefit.

In practice, flexible data utilization makes it possible to (i) use different sampling characteristics in each path and (ii) perform a cross-consistency check to stabilize training. Together with sampling stochasticity, it defines a general design space for splitting-based SSL reconstruction within UNITS.

\section{Experiments}
\subsection{Dataset and Undersampling Masks}
The 2D cardiac Cine dataset used in all experiments is an in-house dataset, which was acquired using a balanced steady-state free precession (bSSFP) sequence on a 1.5T MRI (MAGNETOM Aera, Siemens Healthineers, Erlangen, Germany). The sequence parameters are as follows: TE/TR=1.06/2.12 ms, flip angle=52°, bandwidth=915 Hz/px, spatial resolution=1.9 mm $\times $ 1.9 mm, slice thickness=8 mm, cardiac phases=25. The dataset comprised 95 subjects in total, including 74 patients with cardiovascular disease and 21 healthy subjects. Among them, 82 subjects (65 patients, 17 healthy volunteers) were designated for training, with the remaining subjects used for testing. This study was approved by the local ethics committee (426/2021BO1, 721/2012BO1), and all subjects gave written consent.

The undersampling masks used in all experiments are generated by variable density incoherent spatiotemporal acquisition (VISTA)~\cite{ahmad2015variable}, which can generate spatiotemporal sampling patterns with high levels of uniformity and incoherence while maintaining a constant temporal resolution. Coil sensitivity maps were estimated from the acquired auto-calibration signal using ESPIRiT~\cite{uecker2014espirit} and were compressed to 15 coils using the Berkeley Advanced Reconstruction (BART) toolbox~\cite{uecker2015berkeley}.

\subsection{Implementation Details}
The proposed UNITS framework is agnostic to the network architecture. In this study, the reconstruction network operates in the image domain. We employed a physics-based unrolled neural network with 6 unrolls, each consisting of a UNet regularizer and a data consistency (DC) layer~\cite{xu2025attention}. The encoder and decoder of each UNet contain two stages, in which 2D+t convolutions are performed by applying a 2D spatial convolution followed by a 1D temporal convolution. The spatial and temporal kernel sizes were set to 5 and 3, respectively, and the initial number of convolutional filters was 12. The DC layer is realized via a gradient descent algorithm, and the entire network contains 834,720 trainable parameters.

The model is implemented with complex-valued operations, including complex-valued convolutions~\cite{trabelsi2017deep} and ModReLU activations~\cite{arjovsky2016unitary}. All implementations were conducted in TensorFlow v2.6.0 with Keras v2.6.0, while complex-valued operations were supported by MERLIN v0.3~\cite{hammernik2022machine}. Networks were trained using the Adam optimizer~\cite{kingma2014adam} with a learning rate of $4\times 10^{-4}$ and a batch size of 1. The source code is publicly available: (to be released upon acceptance). 

\subsection{Training Configurations}
\subsubsection{\textit{UNITS-Base}}
In the initial undersampling stage, VISTA masks were applied retrospectively at each training step with random generation seeds and random acceleration rates between $R=2$ and $R=16$. During re-undersampling, two subsets were generated from the acquired points by uniform random selection with a randomly chosen ratio between 0 and 1. One subset was used to construct the input, while the other provided the supervision signal. The two subsets were sampled independently, ensuring diverse input–supervision pairings across training iterations.

\subsubsection{\textit{UNITS-Cross}}
As the extension of \textit{UNITS-Base}, \textit{UNITS-Cross} adopts the same sampling configurations, with the only difference being the use of the cross-consistency loss introduced in Section~\ref{sec:units_cross}.

\subsection{Comparative Experiments}
To evaluate the proposed variants, we compared \textit{UNITS-Base} and \textit{UNITS-Cross} to representative methods SSDU~\cite{yaman2020self} and Noisier2Noise~\cite{millard2023theoretical} under an identical formalism. All experiments used the same dataset with the same reconstruction network to ensure fairness, while preserving the sampling strategies defined in the original works. Subject-specific approaches using only a single scan and generative models with distinct network backbones were therefore excluded. A supervised model trained on fully-sampled images was included as a reference.

\subsection{Ablation Studies}
\subsubsection{Ablation on Sampling Stochasticity}
\begin{table*}[t]
\centering
\caption{Experiment settings of ablation study on sampling stochasticity}
\label{tab:randomness}
\resizebox{\linewidth}{!}{%
\renewcommand{\arraystretch}{1.1}
    \begin{tabular}{cc|cc|ccc}
        \toprule
        \multirow{2}{*}{\textbf{Experiments}} & \multirow{2}{*}{} 
        & \multicolumn{2}{c|}{\textbf{Initial undersampling mask $\mathbf{M_{y}}$}} 
        & \multicolumn{3}{c}{\textbf{Re-undersampling masks $\mathbf{M_{1}}$ and $\mathbf{M_{2}}$}} \\
        \cmidrule(lr){3-4} \cmidrule(lr){5-7} 
        & & \textbf{Generation seed} & \textbf{Acceleration rate} & \textbf{$\mathbf{M_{1}/M_{2}}$ dependence} & \textbf{Ratio of $\mathbf{M_{1}}$ (input)} & \textbf{Ratio of $\mathbf{M_{2}}$ (loss)} \\
        \midrule
        \textit{FixedSampling} 
        & & Fixed & Fixed ($R=8$) & Disjoint & Fixed ($0.4$) & Fixed ($0.6$)\textsuperscript{\dag} \\
        \textit{RandInitSeed} 
        & & Random & Fixed ($R=8$) & Disjoint & Fixed ($0.4$) & Fixed ($0.6$)\textsuperscript{\dag} \\
        \textit{RandRatio} 
        & & Random & Fixed ($R=8$) & Disjoint & Random ($0\sim 1$) & Random ($0\sim 1$)\textsuperscript{\dag} \\
        \textit{IndependentMask} 
        & & Random & Fixed ($R=8$) & Independent & Random ($0\sim 1$) & Random ($0\sim 1$) \\
        \textit{UNITS-Base} 
        & & Random & Random ($R=2\sim16)$ & Independent & Random ($0\sim 1$) & Random ($0\sim 1$) \\
        \bottomrule
    \end{tabular}%
}
\begin{tablenotes}[flushleft]
\scriptsize
\item[*] \textsuperscript{\dag}Note: In the disjoint setting, $M_{2}$ is uniquely determined by $M_{1}$ (i.e., $M_{2}=M_{y}\setminus M_{1}$). Reported ratios of $M_{2}$ therefore reflect the complement of $M_{1}$ rather than an independently selected parameter.
\end{tablenotes}
\end{table*}

To explore how the stochastic sampling introduced in Section~\ref{sec:method_stochasticity} affects reconstruction performance, we designed a series of experiments that progressively incorporate the described stochastic elements. All training configurations are summarized in Table~\ref{tab:randomness}. 

Specifically, we included \textit{FixedSampling} as a deterministic baseline with a fixed initial undersampling mask ($R=8$), a fixed re-undersampling ratio of the input subset ($0.4$), and a disjoint partition of input and loss subsets, similar to SSDU~\cite{yaman2020self}. Starting from this baseline, we incrementally introduced the stochastic elements supported by UNITS: \textit{RandInitSeed} relaxes the constraint of the initial undersampling mask, allowing for random generation seeds at each training step while keeping the acceleration rate constant ($R=8$). \textit{RandRatio} further randomized the re-undersampling ratio, so that the relative sizes of input and loss subsets varied across iterations. \textit{IndependentMask} removed the disjoint constraint, allowing the two subsets to be sampled independently with separate random re-undersampling ratios. Finally, \textit{UNITS-Base} incorporated all of the above and additionally randomized the initial acceleration rate ($R=2\sim 16$). All variants used the same reconstruction network, differing only in their undersampling strategies.

\subsubsection{Ablation on Cross-consistency Loss}
To investigate the effect of the cross-consistency loss, we compared the reconstruction performance of \textit{UNITS-Base} and \textit{UNITS-Cross} under different acceleration factors. Both variants were trained with identical network architectures and undersampling settings, differing only in whether the cross-consistency loss was applied during training.

\subsection{Evaluation Protocol}
\subsubsection{Inference Scenarios}
We evaluated model performance under two inference scenarios: in-distribution (ID) and out-of-distribution (OOD). In the ID setting, the input follows the same procedure as training, meaning the initially undersampled k-space ($R=8$) is further re-undersampled with a ratio ($0.4$). In the OOD setting, the input is directly the initially acquired undersampled k-space without further re-undersampling, which deviates from the training distribution and simulates real-world deployment, where all acquired data are used for reconstruction.

\subsubsection{Evaluation Metrics}
Both quantitative and qualitative evaluations were provided in the results. Quantitative metrics included the mean squared error (MSE), peak signal-to-noise ratio (PSNR), and structural similarity index (SSIM) computed between the reconstructed and fully-sampled images across all test subjects.

\section{Results}
\subsection{Reconstructions of \textit{UNITS-Base}}
\begin{figure*}[t]
    \centering
    \includegraphics[width=\linewidth]{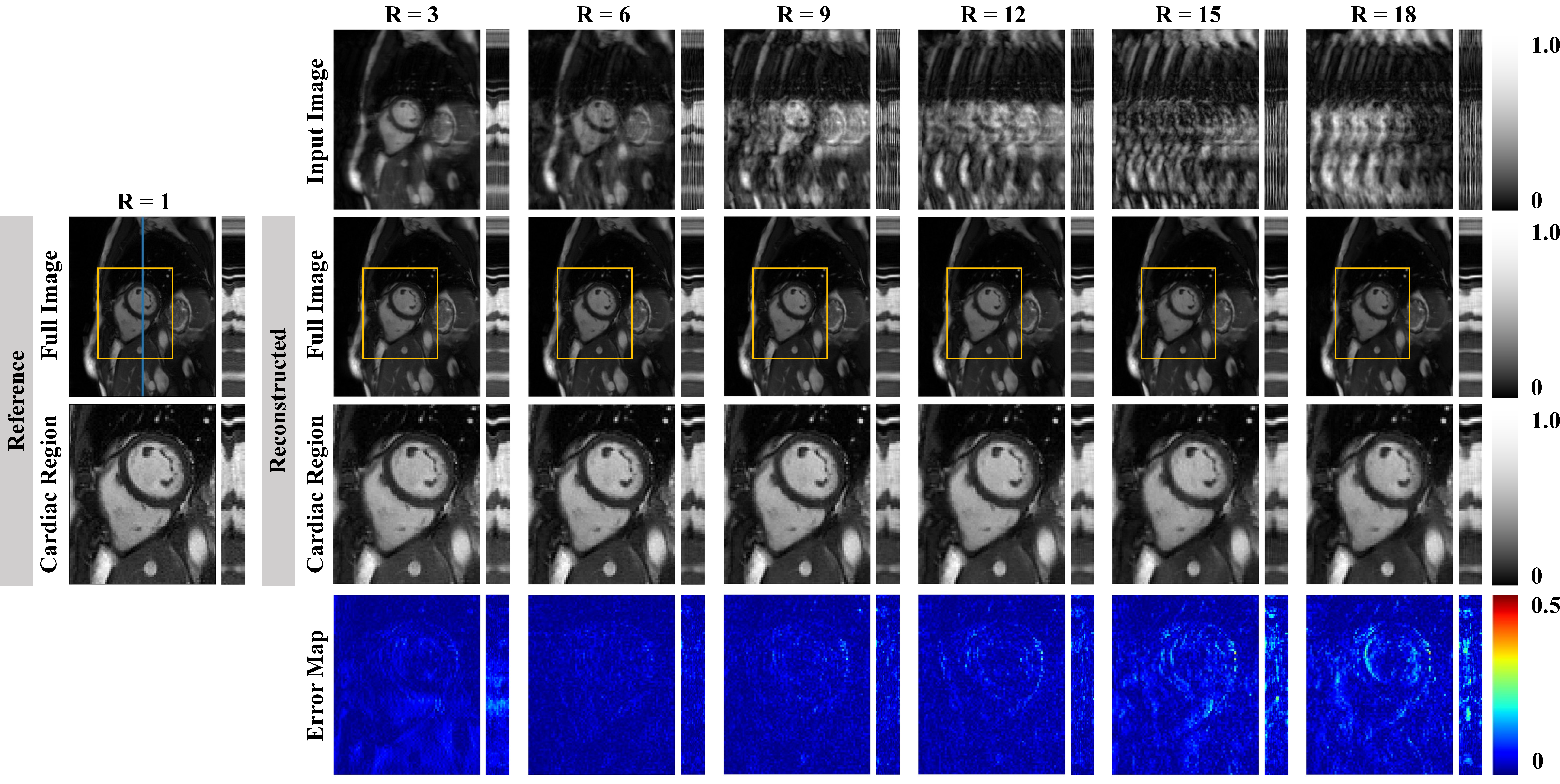}
    \caption{\textbf{Reconstructions in spatial (x-y) and spatiotemporal (y-t) plane of the proposed \textit{UNITS-Base}.} Each column shows the results for the acceleration rates $R=3,6,9,12,15,18$. The first row presents the undersampled zero-filled input images, the second row shows the reconstructed full images, with enlarged cardiac regions (yellow box) displayed in the third row. The bottom row presents the corresponding $2\times$ scaled relative error maps between the reconstructed and the fully-sampled reference. The dynamic performance in the y-t plane corresponds to the blue line in the reference x-y plane image.}
    \label{fig:UNITS_Recon_results}
\end{figure*}

Fig.~\ref{fig:UNITS_Recon_results} shows representative reconstructions obtained with \textit{UNITS-Base}, the basic variant of the proposed framework. We observe that \textit{UNITS-Base} generalizes effectively across all acceleration levels ($R=3\sim 18$) in the test subject. We experienced consistent high-quality reconstruction performance across different noise in both spatial (x-y) and spatiotemporal (y-t) domains, demonstrating strong robustness to shifts in sampling density and further validating the effectiveness of controlled randomness as a means of implicit regularization.

From a clinical perspective, higher acceleration rates directly relate to shorter scan times. In a prospective setting, $R=18$ corresponds to reducing the multi-breath-hold cardiac Cine acquisition ($6$ breath-holds of $16$~s each and $20$~s pause in between) of $196$ s scan time to a single breath-hold of about $6$~s, while still retaining diagnostic fidelity. The ability of \textit{UNITS-Base} to maintain image quality across a wide acceleration spectrum highlights its potential for enabling faster, more reliable, and more patient-friendly MRI examinations.

\subsection{Comparison with Supervised and Existing SSL Methods}
\begin{figure*}[t]
    \centering
    \includegraphics[width=\linewidth]{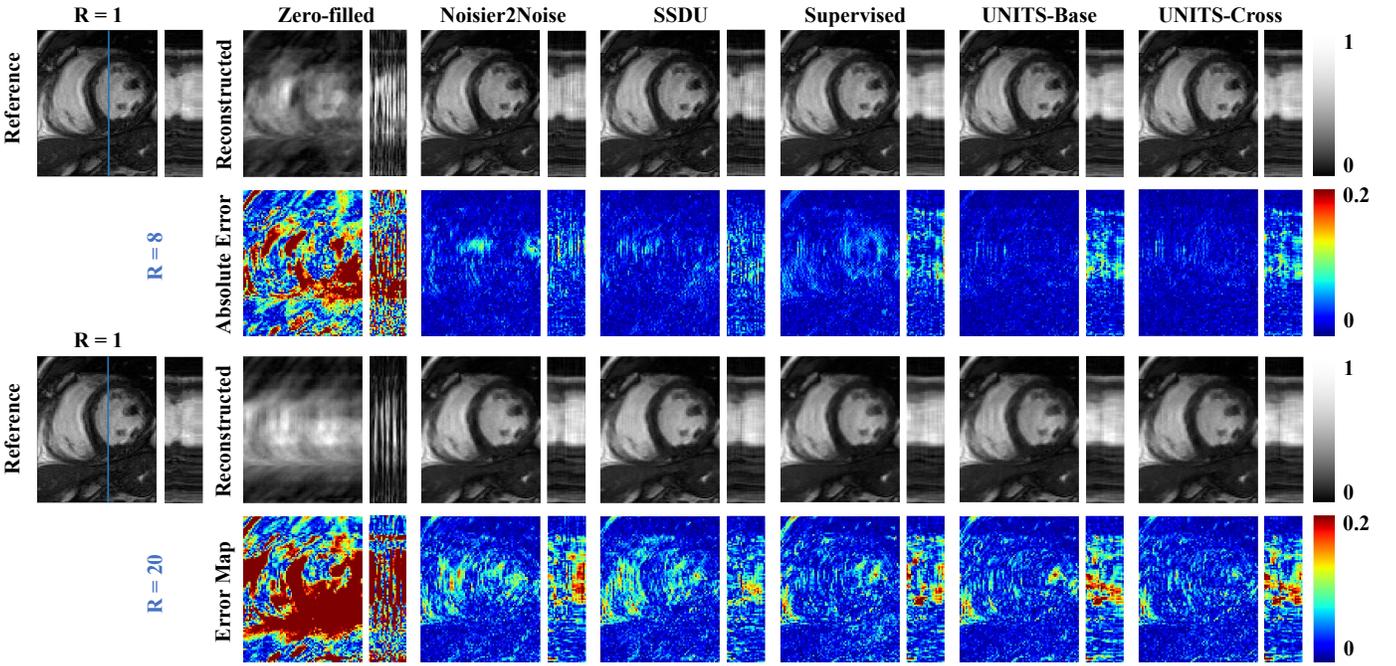}
    \caption{\textbf{Comparison between representative splitting-based self-supervised reconstruction methods within the UNITS framework and supervised learning.} Reconstructions in spatial (x-y) and spatiotemporal (y-t) planes are shown for \textit{zero-filling}, \textit{Noisier2Noise}~\cite{millard2023theoretical}, \textit{SSDU}~\cite{yaman2020self}, supervised learning, \textit{UNITS-Base}, and \textit{UNITS-Cross}. All methods were implemented within the UNITS framework using the same network backbone. Both the initially undersampled k-space ($R=8$, top) and the re-undersampled k-space with ratio $0.4$ (effective acceleration $R=20$, bottom) are evaluated as inference inputs. The dynamic performance in the y-t plane corresponds to the blue line in the reference x-y plane image. The error plots present the corresponding $5\times$ scaled relative error maps between the reconstructed images and the fully-sampled reference.}
    \label{fig:compare_results}
\end{figure*}

Fig.~\ref{fig:compare_results} shows that the evaluated SSL methods achieve reconstruction quality comparable to supervised learning, while \textit{UNITS-Base} and \textit{UNITS-Cross} yield further improvements, particularly in preserving image intensity and reducing residual errors. 

Both UNITS variants can effectively reconstruct undersampled inputs with high image quality comparable to supervised learning. This observation is consistent with the risk-based interpretation in Section~\ref{sec:risk_characterization}, where the self-supervised learning preserves the same pointwise Bayes-optimal target as the corresponding supervised learning under positive supervision weights. Although the analysis is formulated at the population level, the finite-data experiments already exhibit empirical behavior aligned with the theoretical interpretation.

Moreover, \textit{UNITS-Base} and \textit{UNITS-Cross} even present lower residual errors than the supervised baseline in this representative case. We hypothesize that this difference arises from intrinsic biases in the reference images used for supervised training. Specifically, the “fully-sampled” cardiac Cine dataset used for training was acquired in clinical practice with parallel imaging ($2\times$ GRAPPA reconstruction~\cite{griswold2002generalized}). While these images provide sufficient diagnostic quality, they may contain inherent imperfections due to coil sensitivity estimation or interpolation errors. When such reconstructions are used as ground truth, the achievable performance of supervised learning is limited by these biases. In contrast, the self-supervised strategy embodied by UNITS learns directly from the acquired undersampled measurements, thereby avoiding interference from potentially biased reference data.

\subsection{Ablation on Sampling Stochasticity}
\begin{figure*}[t]
\centering
\includegraphics[width=\linewidth]{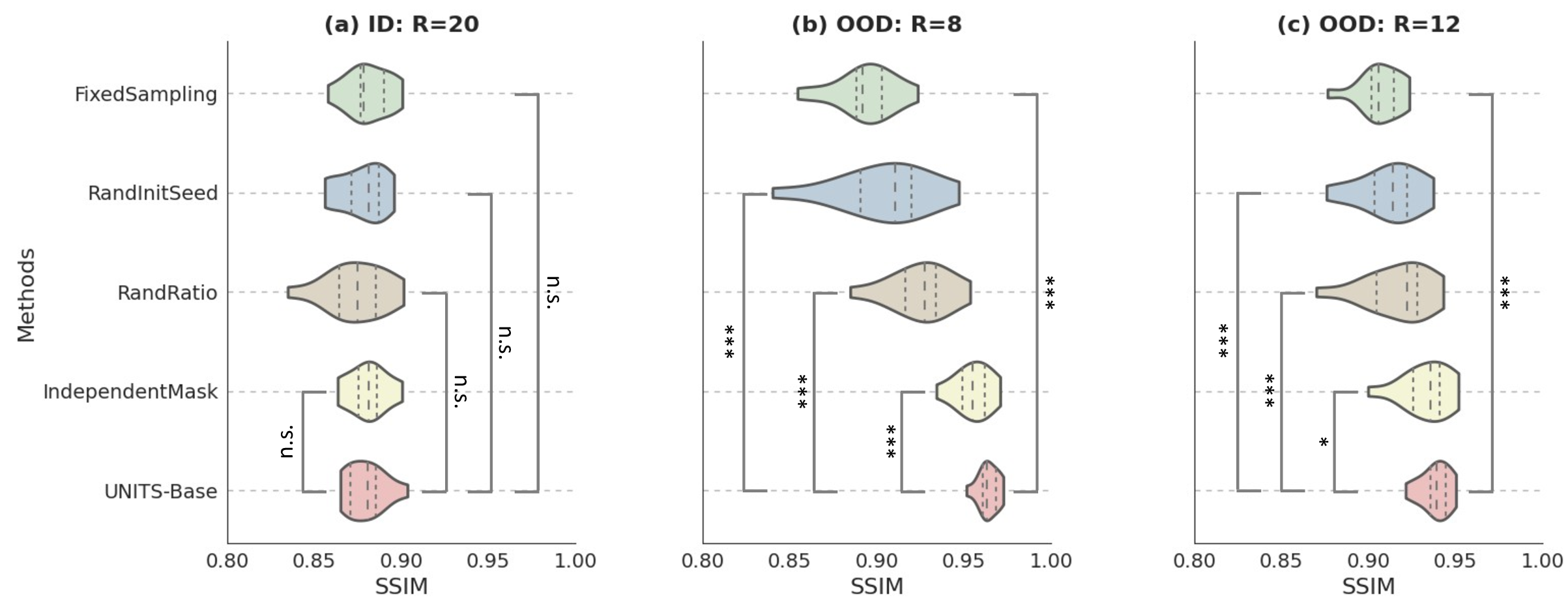}
\caption{\textbf{Ablation results on sampling stochasticity.} Quantitative comparison of the experiments summarized in Table~\ref{tab:randomness}, evaluated using structural similarity index (SSIM) across all slices of all test subjects under three inference conditions: (a) in-distribution (ID): the input is re-undersampled from an initially undersampled k-space ($R=8$) with ratio $0.4$, yielding an effective acceleration of $R=20$ (matching the training setup of \textit{FixedSampling}). (b,c) out-of-distribution (OOD): the input is the initially undersampled k-space with acceleration (b)$R=8$ and (c)$R=12$, without further re-undersampling. Violin plots depict the SSIM distribution, with vertical dashed lines indicating the median and interquartile ranges. Asterisks denote statistically significant differences assessed by the Wilcoxon signed-rank test across subjects (\text{*}: $p < 0.05$; \text{*}: $p < 0.01$; \text{*}: $p < 0.001$; n.s.: not significant).}
\label{fig:Result_Randomness}
\end{figure*}

Fig.~\ref{fig:Result_Randomness} shows the SSIM values of reconstructions obtained from \textit{FixedSampling} to \textit{UNITS-Base}, demonstrating the effect of progressively increased sampling stochasticity under both ID and OOD inference scenarios. We discovered that the deterministic baseline, \textit{FixedSampling}, achieves performance comparable to the stochastic variants when the test data distribution exactly matches the training distribution (Fig.~\ref{fig:Result_Randomness}(a)). However, its performance drops and becomes the worst under OOD conditions, indicating its lack of robustness to sampling variability. 

As increasing levels of stochasticity are introduced (from \textit{FixedSampling} to \textit{UNITS-Base}), the models exhibit progressively improved generalization, with higher SSIM scores and reduced variance shown in Fig.~\ref{fig:Result_Randomness}(b) and (c). Among them, \textit{UNITS-Base}, which integrates all stochastic enhancement strategies, delivers the most consistent and robust performance under OOD scenarios. Together, these findings confirm that sampling stochasticity is key to bridging the training–inference distribution gap.

\subsection{Ablation on Cross-consistency Loss}
\begin{table}[t]
\caption{\textbf{Quantitative evaluation of \textit{UNITS-Base} and \textit{UNITS-Cross}}: Reported are the average and standard deviation of mean squared error (MSE), peak signal-to-noise ratio (PSNR) in dB, and structural similarity index (SSIM) of reconstructed images compared to fully-sampled references, across all test subjects under acceleration factors $R=8$, $R=12$, and $R=16$ (mean±std). The best performance metrics are indicated in bold.}
\label{tab:ablation_quant}
\resizebox{\linewidth}{!}{%
\renewcommand{\arraystretch}{1.0}
    \begin{tabular}{cc@{}|c@{\hspace{3pt}}c|c@{\hspace{3pt}}c|c@{\hspace{3pt}}c}
        \toprule
        \multirow{2}{*}{\textbf{Metrics}} & \multirow{2}{*}{} 
        & \multicolumn{2}{c|}{$R=8$} 
        & \multicolumn{2}{c|}{$R=12$}
        & \multicolumn{2}{c}{$R=16$} \\
        \cmidrule(lr){3-4} \cmidrule(lr){5-6} \cmidrule(lr){7-8}
        & & \textit{UNITS-Base} & \textit{UNITS-Cross} & \textit{UNITS-Base} & \textit{UNITS-Cross} & \textit{UNITS-Base} & \textit{UNITS-Cross} \\
        \midrule
        \textbf{MSE} 
        & & 4.16 ± 1.57 & \textbf{3.79 ± 1.31} & 8.68 ± 3.58 & \textbf{7.73 ± 3.49} & 14.71 ± 7.47 & \textbf{13.49 ± 7.36} \\
        \textbf{PSNR}
        & & 38.08 ± 0.90 & \textbf{38.08 ± 0.90} & 34.76 ± 0.87 & \textbf{35.18 ± 0.79} & 32.54 ± 0.88 & \textbf{33.04 ± 0.88} \\
        \textbf{SSIM} 
        & & 0.96 ± 0.01 & \textbf{0.97 ± 0.01} & \textbf{0.94 ± 0.01} & \textbf{0.94 ± 0.01} & 0.91 ± 0.01 & \textbf{0.92 ± 0.01} \\
        \bottomrule
    \end{tabular}%
}
\end{table}

Quantitative comparisons between \textit{UNITS-Base} and \textit{UNITS-Cross} under three different acceleration rates ($R=8,12,16$) are summarized in Table~\ref{tab:ablation_quant}. Across all accelerations, \textit{UNITS-Cross} consistently achieves lower MSE, higher PSNR, and higher SSIM than \textit{UNITS-Base}. While the absolute differences are modest, the systematic trend indicates more stable reconstruction quality. 

The variability across test subjects is also slightly reduced for \textit{UNITS-Cross}, suggesting more consistent reconstruction quality across the test cohort. This empirical trend is compatible with the motivation of cross-consistency in Section~\ref{sec:units_cross}. Overall, \textit{UNITS-Cross} provides a more stable training strategy and makes more effective use of the available data, illustrating how UNITS can flexibly leverage multiple sampling realizations without changing the reconstruction architecture.

\section{Discussion}
In this study, we introduced UNITS as a general framework for splitting-based self-supervised MRI reconstruction. Conceptually, UNITS can describe a broader set of existing methods through a unified formalism rather than treating them as method-specific implementations, thereby providing a consistent framework for comparing data-flow and design choices. This generalizability stems from two key concepts of UNITS: sampling stochasticity and flexible data utilization, through which different methods can be viewed as specific instantiations of UNITS by specifying the sampling choices and the assignment of subsets.

Theoretically, we characterized the population risk of self-supervised learning and found that the self-supervised risk can be written as a sampling-weighted supervised risk under pointwise loss. As a result, when the induced supervision weights are positive, the pointwise Bayes-optimal predictor of UNITS is the same as supervised learning. These analyses provide a principled explanation for the empirical success of splitting-based self-supervision.


\subsection{Theoretical Boundary and Practical Implications}
The pointwise equivalence of the Bayes-optimal predictor in Proposition~\ref{proposition:equivalence} relies on the positivity of the supervision weight $w_i(Y_1)$. Through the explicit form of $w_i(Y_1)$ in Proposition~\ref{proposition:wi(Y1)}, this condition can be expressed as concrete constraints on the sampling probabilities: each k-space location must have a non-zero possibility of being acquired initially ($p_i>0$), must not be deterministically included in the input subset ($q_i<1$), and must have a non-zero probability of being selected for supervision ($r_i>0$).

In practice, the re-undersampling probabilities $q_i$ and $r_i$ are usually straightforward to enforce, since re-undersampling is performed retrospectively during training. The initial sampling probability $p_i$ is more restrictive, especially for prospective acquisition with a fixed undersampling mask. In this case, some k-space locations may never be observed ($p_i=0$), and their pointwise Bayes-optimal predictor can not be learned directly from the self-supervised loss.

Nevertheless, training with fixed initial masks can still work well in practice, as observed in the \textit{FixedSampling} baseline. The reason is that MRI reconstruction is not a set of isolated pointwise prediction tasks. MR images are highly structured, and Fourier encoding couples local image structures with global k-space measurements. Reconstruction networks further exploit these dependencies through convolutional receptive fields. Consequently, locations without direct pointwise supervision may still be estimated indirectly from consistency constraints on the acquired samples. Therefore, the UNITS theory identifies the conditions required for direct pointwise guarantees, whereas the empirical success of fixed initial masks reflects a practical relaxation enabled by image priors.

\subsection{Design Flexibility of UNITS}
Beyond its theoretical contributions, UNITS identifies two flexible design dimensions: sampling stochasticity and flexible data utilization. Related strategies, such as random mask generation, repeated partitioning, and disjoint or independent splitting, have appeared in some earlier works. The role of UNITS is to make these choices explicit within a common framework. UNITS enables different training pipelines to be described consistently and provides a basis for analyzing how mask distributions, re-undersampling ratios, and subset dependence influence the training behavior. From these two concepts stems our benchmark variants: \textit{UNITS-Base} and \textit{UNITS-Cross}. 

\textit{UNITS-Base} uses sampling stochasticity by allowing for variable acceleration factors, random re-undersampling ratios, and independently generated re-undersampling masks. The enhanced sampling variability acts as an implicit regularization and improves robustness to distribution shifts at inference. Such robustness is crucial in clinical practice, where undersampling patterns and acceleration rates often vary across subjects, sequences, and acquisition protocols. It also addresses a common training-inference mismatch in splitting-based SSL: models are trained on further re-undersampled data, but are often tested on the initially acquired measurements. By exposing the network to a wider range of sampling conditions during training, \textit{UNITS-Base} reduces the performance degradation caused by this mismatch.

Flexible data utilization motivates the cross-consistency loss used in \textit{UNITS-Cross}. By enforcing consistency across independently sampled k-space subsets, \textit{UNITS-Cross} exploits more of the available measurement information without changing the reconstruction architecture. In our experiments, \textit{UNITS-Cross} consistently improves over \textit{UNITS-Base} across acceleration rates and shows slightly lower variability across test subjects. The variability reported in Table~\ref{tab:ablation_quant} reflects inter-subject differences in reconstruction metrics, whereas Appendix~E analyzes the variance of the prediction bias at a fixed k-space location. These quantities are not identical, but the empirical trend is consistent with the interpretation that cross-consistency can stabilize reconstruction by leveraging complementary supervision from multiple sampling realizations. We also observed faster convergence for \textit{UNITS-Cross} than for its single-loss counterpart and other SSL baselines, suggesting more efficient use of the available measurements.

\subsection{Relation to Existing Theory}
The Noisier2Noise-based analysis of Millard and Chiew~\cite{millard2023theoretical} provides an important theoretical work for SSL reconstruction. Their study extends Noisier2Noise to variable-density undersampled MRI and interprets SSDU as a version of Noisier2Noise with a particular rank-deficient loss weighting, under which SSDU is shown to be theoretically correct in expectation for an $\ell_2$ objective.

Our UNITS work differs primarily in the scope of its formulation. Noisier2Noise models self-supervision as learning the initially undersampled data from a further re-undersampled input. In contrast, UNITS formulates self-supervision directly as learning between re-undersampled subsets. This subset-to-subset view does not fix the supervision to the initially acquired k-space, and therefore allows different supervision designs to be described within the same framework. For example, Noisier2Noise-like supervision corresponds to using an identity supervision mask, whereas SSDU-like training corresponds to a particular disjoint subset assignment.

This broader formulation also changes the theoretical analysis. Millard and Chiew focused on expectation consistency under a squared-error loss. UNITS instead characterizes the self-supervised risk under arbitrary pointwise losses and expresses it as a sampling-weighted supervised risk. The induced supervision weights are given explicitly by the re-undersampling mechanism, which makes the role of the sampling design transparent. Under the $\ell_2$ loss, the same weighting further relates the training residual to the prediction bias. As such, UNITS generalizes method-specific theoretical interpretations into a unified risk characterization for splitting-based self-supervision.

\subsection{Relation to Existing SSL Methods}
UNITS is designed for splitting-based self-supervised MRI reconstruction approaches, where acquired measurements are divided into subsets that serve as network inputs or supervision signals. Within this scope, different methods can be described by specifying the undersampling mechanism, the number and assignment of re-undersampled subsets, the dependence between subsets, and the reconstruction backbone. The theoretical results apply when the training objective follows the splitting-based loss considered in Section~\ref{sec:risk_characterization} and satisfies the corresponding sampling conditions.

Representative examples are shown in the supplementary figures. SSDU (Fig.~S3) corresponds to the case of a fixed initial mask, two strictly complementary subsets, and a deterministic re-undersampling strategy. ZS-SSL (Fig.~S4) follows the same measurement-splitting principle in a subject-specific setting, with an additional validation subset for early stopping. SSDiffRecon (Fig.~S5) also adopts a two-subset splitting strategy, but replaces the unrolled reconstruction network with a diffusion-based denoiser. 

Therefore, UNITS enhances the interpretability of earlier self-supervised methods, many of which were developed empirically or heuristically, and further establishes UNITS as a standardized benchmark for systematic comparison across reconstruction strategies.

\subsection{Limitations and Future Work}
UNITS focuses on splitting-based self-supervision methods. Methods that acquire the same subject multiple times, such as Noise2Noise~\cite{lehtinen2018noise2noise} and RARE~\cite{liu2020rare}, are closely related but fall outside the present formulation, since their supervision comes from independent acquisitions rather than re-undersampled subsets of a single acquisition. Extending UNITS to this setting would require additional modeling of inter-scan consistency, motion, phase variation, and measurement noise.

The bias analysis shows how sampling probabilities and subset dependence affect the training residual, but how these probabilities should be chosen most efficiently in practice continues to be an area of active investigation. Independent re-undersampling increases diversity and retains additional residual pathways, but it may also introduce overlap or leave some acquired samples unused in a given iteration. Repeated stochastic sampling may compensate for these omissions over training, yet a precise characterization of this trade-off is still lacking.

We demonstrated UNITS on cardiac Cine MRI to validate the framework. Further validation across anatomies, contrasts, trajectories, and prospective acquisitions will be studied to validate the generality.


\section{Conclusion}
In this work, we introduced UNITS, a unified framework for splitting-based self-supervised MRI reconstruction. By formulating training as learning between re-undersampled subsets, UNITS provides a common language for describing existing splitting-based methods and for designing new variants. Theoretically, we showed that the self-supervised risk under pointwise losses can be expressed as a sampling-weighted supervised risk. Under positive supervision weights, this weighted risk preserves the same pointwise Bayes-optimal predictor as the corresponding supervised learning. UNITS further links the training residual to the prediction bias, revealing how re-undersampling mechanisms influence the effective training signal.

Building on this theoretical foundation, the framework remains flexible in practice through sampling stochasticity and flexible data utilization, which are instantiated in \textit{UNITS-Base} and \textit{ UNITS-Cross}. Experiments on cardiac Cine MRI show that these variants achieve reconstruction quality comparable to supervised learning, improve robustness to sampling shifts, and benefit from complementary supervision without architectural changes. Together, these results establish UNITS as a theoretical and practical framework for interpretable splitting-based self-supervised reconstruction, and provide a basis for future work on sampling strategy design.

\section*{Appendix}
To facilitate understanding of the derivation, Table~\ref{tab:notation} summarizes the symbols and variables used throughout the proof.

\begin{table*}[t]
\centering
\caption{Summary of notation and variable definitions}
\label{tab:notation}
\begin{tabular}{lll}
\toprule
\textbf{Symbol}  & \textbf{Definition} & \textbf{Description} \\
\midrule
$e$           & Prediction bias                           & $e(Y_1):=f(Y_1)-f^{\star}(Y_1)$\\
$f$           & Reconstruction network                    & $f: \mathbb{C}^{N}\to \mathbb{C}^{N}$ \\
$J(f;Y_1)$    & Conditional risk term                     & $J(f;Y_1):=\mathbb E\!\left[M_{Y_2}\,\ell(f(Y_1),Y_{0,i})\;\middle|\; Y_1\right]$\\
$k_{i}$       & Conditional probability of location $i$   & $k_{i}:=P[Y_{i}= 0\mid Y_{1,i}= 0]$\\
$\ell$        & Pointwise loss function                   & $\ell:\mathbb{C}\times \mathbb{C}\to\mathbb{R}_{+}$ \\
$M_Y$         & Initial undersampling mask                & $M_{Y}\in\left \{ 0,1 \right \}^{N}$ \\
$M_1$         & Re-undersampling mask                     & $M_1\in\left \{ 0,1 \right \}^{N}$\\
$M_2$         & Re-undersampling mask                     & $M_2\in\left \{ 0,1 \right \}^{N}$\\
$M_{Y_{1}}$   & Effective sampling mask of $Y_1$          & $M_{Y_{1}}=M_{1}\odot M_{Y}$\\
$M_{Y_{2}}$   & Effective sampling mask of $Y_2$          & $M_{Y_{2}}=M_{2}\odot M_{Y}$\\
$p_{i}$       & Initial undersampling probability of location $i$       & $P[M_{y, ij}=1]$ \\
$q_{i}$       & Re-undersampling conditional probability of $M_{1,i}$   & $P[Y_{1,i}\neq 0\mid Y_{i}\neq 0]$ \\
$r_{i}$       & Re-undersampling conditional probability of $M_{2,i}$   & $P[Y_{2,i}\neq 0\mid Y_{i}\neq 0]$ \\
$w_i(Y_1)$    & conditional supervision weight of location $i$          & $w_i(Y_1):=\mathbb{E}[M_{{Y_2,i}}\mid Y_1]$\\
$Y_0$         & Fully-sampled k-space                     & $Y_0\in \mathbb{C}^{N}$ \\
$Y$           & Initial undersampled k-space              & $Y=M_{y}\odot Y_{0}$ \\
$Y_1$         & Re-undersampled k-space                   & $Y_1 = M_1\odot Y$ \\
$Y_2$         & Re-undersampled k-space                   & $Y_2 = M_2\odot Y$ \\
\bottomrule
\end{tabular}
\end{table*}

\subsection{Proof of Theorem~\ref{thm:equivalence}}
Given the pointwise loss Eq.~\eqref{eq:pointwise_loss_func} and using $Y_2=M_{Y_{2}}\odot Y_{0}$, the self-supervised risk in Eq.~\eqref{eq:risk_ssl} can be written as:
\begin{equation}
    \mathcal R_{\mathrm{ssl}}(f)=\mathbb E\!\left[\sum_{i=1}^N M_{Y_2,i}\,\ell(f_i(Y_1),Y_{0,i})\right].
\end{equation}
Applying the tower property with respect to $Y_1$ yields:
\begin{equation}
\begin{split}
    \mathcal R_{\mathrm{ssl}}(f)&=\mathbb E\!\left[\mathbb E\!\left[\sum_{i=1}^N M_{Y_2,i}\,\ell(f_i(Y_1),Y_{0,i})\;\middle|\; Y_1\right]\right]\\
    &=\mathbb E\!\left[\sum_{i=1}^N\mathbb E\!\left[M_{Y_2,i}\,\ell(f_i(Y_1),Y_{0,i})\;\middle|\; Y_1\right]\right]\\
    &=\mathbb E\!\left[\sum_{i=1}^N J_i(f;Y_1)\right],
    \label{eq:pointwise_ssl_risk}
\end{split}
\end{equation}
where $J_i(f;Y_1):=\mathbb E\!\left[M_{Y_2,i}\,\ell(f_i(Y_1),Y_{0,i})\;\middle|\; Y_1\right]$ denotes the $i$-th conditional risk term.

We now focus on a particular location $i$, where $1\le i\le N$ and $i\in \mathbb{Z}$. Given the network input $Y_{1}$, the contribution of this location to the loss depends on whether it is included in the supervision subset. Under the UNITS framework, this inclusion is determined by the re-undersampling mechanism and does not further depend on the unknown ground truth $Y_0$. Hence:
\begin{equation}
    \mathbb E[M_{Y_2,i}\mid Y_1,Y_0]=\mathbb E[M_{Y_2,i}\mid Y_1].
\end{equation}
We therefore define the re-undersampling induced weight as:
\begin{equation}
    w_i(Y_1):=\mathbb E[M_{Y_2,i}\mid Y_1].
\end{equation}
Since $f_i(Y_1)$ is measurable with respect to $Y_1$, we obtain:
\begin{equation}
\begin{split}
    J_i(f;Y_1)&=\mathbb{E}[\mathbb{E}[M_{Y_2,i}\,\ell(f_i(Y_1),Y_{0,i})|\; Y_1,Y_{0}]|\; Y_1]\\
&=\mathbb{E}[\ell(f_i(Y_1),Y_{0,i})\mathbb{E}[M_{Y_2,i}\,|\; Y_1,Y_{0}]|\; Y_1]\\
&=w_i(Y_1)\,\mathbb E\!\left[\ell(f_i(Y_1),Y_{0,i})\;\middle|\; Y_1\right].
\end{split}
\end{equation}
Substituting back to Eq.~\eqref{eq:pointwise_ssl_risk} gives:
\begin{equation}
    \mathcal R_{\mathrm{ssl}}(f)=\mathbb E\!\left[\sum_{i=1}^Nw_i(Y_1)\,\mathbb E\!\left[\ell(f_i(Y_1),Y_{0,i})\;\middle|\; Y_1\right]\right].
\end{equation}
This proves the claim.
\hfill $\blacksquare$

\subsection{Proof of Proposition~\ref{proposition:wi(Y1)}}
\label{sec:derivation_wi}
The derivation below is inspired by the case-splitting analysis of Millard et al.~\cite{millard2023theoretical}, but is adapted to a different objective and sampling structure. We begin from the definition of $w_i(Y_1)$ in Theorem~\ref{thm:equivalence}:
\begin{equation}
    w_i(Y_1):=\mathbb{E}[M_{{Y_2},i}\mid Y_1]=\mathbb{E}[M_{2,i}M_{Y,i}\mid Y_1].
    \label{eq:wi_definition}
\end{equation}
Since this quantity depends on $Y_1$, we distinguish two cases according to whether the corresponding entry $Y_{1, i}$ is observed.

\begin{figure}[t]
    \centering
    \includegraphics[width=\linewidth]{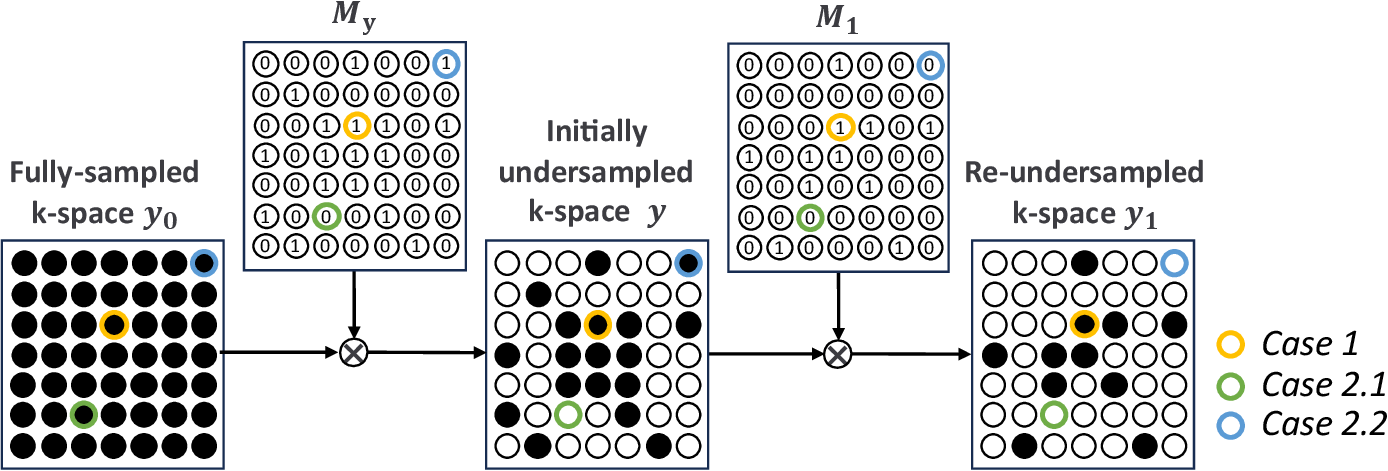}
    \caption{\textbf{Illustration of the case analysis in the proof of Theorem~\ref{thm:equivalence}} In the k-space plots ($y_{0}$, $y$, $y_{1}$), black dots indicate sampled k-space locations and white dots indicate unsampled points. In the mask plots ($M_{y}$ and $M_{1}$), "1" and "0" denote whether a location is sampled or not in the binary mask. Three representative k-space locations are highlighted: \textit{Case 1} (yellow): the location is sampled in both initial and re-undersampling stages; \textit{Case 2.1} (green): the location is not sampled in the initial undersampling; \textit{Case 2.2} (blue): the location is sampled in the initial undersampling but not sampled in the re-undersampling stage.}
    \label{fig:case_analysis}
\end{figure}

\textbf{\textit{Case 1:}} $Y_{1,i}$ is sampled (i.e., $M_{Y,i}M_{1,i}=1$, yellow dot in Fig.~\ref{fig:case_analysis}). Since $Y_{1,i}$ is observed, the corresponding sampling masks $M_{Y,i}$ and $M_{1,i}$ must be one ($M_{Y,i}=M_{1,i}=1$). Consequently, Eq.~\eqref{eq:wi_definition} is simplified to: 
\begin{equation}
    w_i(Y_1)=\mathbb{E}[M_{2,i}\mid Y_{1,i}\ne 0, Y_{1,-i}].
    \label{eq:case.1}
\end{equation}
Here, we denote $Y_{1,-i}$ as the collection of all elements of $Y_{1}$ except for the $i$-th component, i.e., $Y_{1,-i}={Y_{1,j}:j\in \left \{ 1,2,...,N \right \}\setminus \left \{ i \right \}}$.

\textbf{\textit{Case 2:}} $Y_{1,i}$ is not sampled (i.e., $M_{Y,i}M_{1,i}=0$). This can occur either because the location was not sampled in the initial acquisition or because it was sampled initially but excluded during re-undersampling. We therefore further distinguish between whether the corresponding k-space location $Y_{i}$ was sampled or not:

\textbf{\textit{(Case 2.1)}} $Y_{i}$ is not sampled (i.e., $M_{Y,i}=0$, green dot in Fig.~\ref{fig:case_analysis}). Under this case, Eq.~\eqref{eq:wi_definition} is zero:
\begin{equation}
    w_i(Y_1)=0.
\end{equation}

\textbf{\textit{(Case 2.2)}} $Y_{i}$ is sampled (i.e., $M_{Y,i}=1$, blue dot in Fig.~\ref{fig:case_analysis}). Similar to Case 1, Eq.~\eqref{eq:wi_definition} can be simplified as:
\begin{equation}
    w_i(Y_1)=\mathbb{E}[M_{2,i}\mid Y_{1,i}=0, Y_{1,-i},Y_{i}\ne 0].
\end{equation}

By the law of total expectation, the expression of $w_i$ under \textit{Case 2} is obtained by summing the conditional expectations from \textit{Case 2.1} and \textit{Case 2.2}, each weighted by its respective conditional probability:
\begin{equation}
    \resizebox{\linewidth}{!}{$
    \begin{split}
    w_i(Y_1)&=P[Y_{i}=0\mid Y_{1,i}=0,Y_{1,-i}]\cdot w_i(Y_{1,i}=0,Y_{1,-i},Y_{i}=0)\\
    &+P[Y_{i}\ne 0\mid Y_{1,i}=0,Y_{1,-i}]\cdot w_i(Y_{1,i}=0,Y_{1,-i},Y_{i}\ne 0)\\
    &=k_{i}\cdot 0 +(1-k_{i})\cdot \mathbb{E}[M_{2,i}\mid Y_{1,i}=0, Y_{1,-i},Y_{i}\ne 0]\\
    &=(1-k_{i})\cdot \mathbb{E}[M_{2,i}\mid Y_{1,i}=0, Y_{1,-i},Y_{i}\ne 0],
    \end{split}$}
    \label{eq:case.2}
\end{equation}
where
\begin{equation}
    \resizebox{\linewidth}{!}{$
    \begin{split}
    k_{i}&=P\left [Y_{i}=0\mid Y_{1,i}=0 \right ]=\frac{P[Y_{i}=0,Y_{1,i}=0]}{P[Y_{1,i}=0]}\\
    &=\frac{P[Y_{i}=0,Y_{1,i}=0]}{P[Y_{1,i}=0\mid Y_{i}=0]\cdot P[Y_{i}=0]+P[Y_{1,i}=0\mid Y_{i}\ne 0]\cdot P[Y_{i}\ne 0]}\\
    &=\frac{1-p_{i}}{1\cdot (1-p_{i})+(1-q_{i})\cdot p_{i}}=\frac{1-p_{i}}{1-p_{i}q_{i}},    
    \end{split}$}
\label{eq:kj}
\end{equation}
where $P$ denotes the probability measure. To ensure $k_{i}$ is well-defined, we require $1-p_{i}q_{i}>0$.

By combining \textbf{\textit{Case 1}} and \textbf{\textit{Case 2}}, we obtain the following unified expression, which holds for both Eq.~\eqref{eq:case.1} and Eq.~\eqref{eq:case.2}:
\begin{equation}
\resizebox{\linewidth}{!}{$
    w_{i}(Y_{1})=(1-k_i(1-\mathbb{E}[M_{Y,i}M_{1,i}\mid Y_{1}])) \mathbb{E}[M_{2,i}\mid Y_{1},Y_{i}\ne0].
    $}
    \label{eq:both_case_solution}
\end{equation}
Eq.~\eqref{eq:both_case_solution} can be simplified to Eq.~\eqref{eq:case.1} when $Y_{1,i}\ne 0$, in which case $\mathbb{E}[M_{Y,i}M_{1,i}\mid Y_{1}]=1$ and the event $Y_{i}\ne 0$ holds automatically. Likewise, when $Y_{1,i}= 0$, we have $\mathbb{E}[M_{Y,i}M_{1,i}\mid Y_{1}]=0$, and Eq.~\eqref{eq:both_case_solution} reduces to Eq.~\eqref{eq:case.2}.

Substituting Eq.~\eqref{eq:kj} into Eq.~\eqref{eq:both_case_solution} yields the explicit form of $w_i(Y_1)$ in Proposition~\ref{proposition:wi(Y1)}:
\begin{equation}
\resizebox{\linewidth}{!}{$
    w_{i}(Y_{1})=(1-\frac{1-p_{i}}{1-p_{i}q_{i}}(1-\mathbb{E}[M_{Y,i}M_{1,i}\mid Y_{1}])) \mathbb{E}[M_{2,i}\mid Y_{1}, Y_{i}\ne 0].
$}
\end{equation}



\subsection{Proof of Proposition~\ref{proposition:feactorization}}
Since $Y_{2,i}=M_{Y_{2},i}Y_{0,i}$, Eq.~\eqref{eq:training_residual} can be written as:
\begin{equation}
    \mathbb{E}[\mathcal{E}_{i}\mid Y_{1}]=\mathbb{E}[M_{Y_{2},i}\cdot (f_i(Y_{1})-Y_{0,i}) \mid Y_{1}].
    \label{eq:training_residual_2}
\end{equation}
We next relate this training residual to the Bayes-optimal predictor $f^{\star}_i(Y_1)$. By adding and subtracting $f^{\star}_i(Y_1)$ inside the parentheses, we obtain:
\begin{equation}
    \resizebox{\linewidth}{!}{$
    \begin{split}
        \mathbb{E}[\mathcal{E}_{i}\mid Y_{1}]&=\mathbb{E}[M_{Y_{2},i}\cdot (f_i(Y_{1})-f^{\star}_i(Y_1)+f^{\star}_i(Y_1)-Y_{0,i}) \mid Y_{1}]\\
        &=\mathbb{E}[M_{Y_{2},i}\cdot (f_i(Y_{1})-f^{\star}_i(Y_1)) \mid Y_{1}]+\mathbb{E}[M_{Y_{2},i}\cdot (f^{\star}_i(Y_1)-Y_{0,i}) \mid Y_{1}].
        \label{eq:training_residual_3}
    \end{split}$}
\end{equation}
Recall that the prediction bias is defined as $e_{i}(Y_1):=f_i(Y_1)-f^{\star}_i(Y_1)$, and that the conditional supervision weight is given by $w_i(Y_1)=\mathbb{E}[M_{Y_{2},i}\mid Y_{1}]$. Since $e_{i}(Y_1)$ is measurable with respect to $Y_{1}$, the first term in Eq.~\eqref{eq:training_residual_3} simplifies to:
\begin{equation}
    \begin{split}
        \mathbb{E}[M_{Y_{2},i}\cdot (f_i(Y_{1})-f^{\star}_i(Y_1)) \mid Y_{1}]&=\mathbb{E}[M_{Y_{2},i}\cdot e_{i}(Y_1)\mid Y_{1}]\\
        &=\mathbb{E}[M_{Y_{2},i}\mid Y_{1}]\cdot e_{i}(Y_1)\\
        &=w_i(Y_1)\cdot e_{i}(Y_1).
    \end{split}
    \label{eq:residual_first_term}
\end{equation}
Hence, Eq.~\eqref{eq:training_residual_3} becomes:
\begin{equation}
    \mathbb{E}[\mathcal{E}_{i}\mid Y_{1}]=w_i(Y_1)\cdot e_{i}(Y_1)+\mathbb{E}[M_{Y_{2},i}\cdot (f^{\star}_i(Y_1)-Y_{0,i}) \mid Y_{1}].
    \label{eq:training_residual_4}
\end{equation}

This expression makes explicit how the training residual depends on the prediction bias, up to an additional term involving the Bayes-optimal predictor. A further simplification is obtained under the $l_2$ loss, for which Corollary~\ref{corollary:mse} gives:
\begin{equation}
    f^{\star}_i(Y_1)=\mathbb{E}[Y_{0,i}\mid Y_1].
\end{equation}
Substituting it into Eq.~\eqref{eq:training_residual_4} yields:
\begin{equation}
    \mathbb{E}[\mathcal{E}_{i}\mid Y_{1}]=w_i(Y_1)\cdot e_{i}(Y_1)+\mathbb{E}[M_{Y_{2},i}\cdot (\mathbb{E}[Y_{0,i}\mid Y_1]-Y_{0,i}) \mid Y_{1}].
    \label{eq:training_residual_5}
\end{equation}
Under the UNITS framework, the effective supervision mask $M_{Y_{2},i}$ is determined by the sampling process and is independent of the unknown target $Y_{0,i}$ once $Y_1$ is given. Hence, the second term in Eq.~\eqref{eq:training_residual_5} becomes:
\begin{equation}
\resizebox{\linewidth}{!}{$
    \mathbb{E}[M_{Y_{2},i}\cdot (\mathbb{E}[Y_{0,i}\mid Y_1]-Y_{0,i}) \mid Y_{1}]=\mathbb{E}[M_{Y_{2},i}\mid Y_{1}]\cdot \mathbb{E}[\mathbb{E}[Y_{0,i}\mid Y_1]-Y_{0,i} \mid Y_{1}],
$}
\end{equation}
which is zero because:
\begin{equation}
\resizebox{\linewidth}{!}{$
\begin{split}
    \mathbb{E}[\mathbb{E}[Y_{0,i}\mid Y_1]-Y_{0,i} \mid Y_{1}]&=\mathbb{E}[\mathbb{E}[Y_{0,i}\mid Y_1]\mid Y_{1}]-\mathbb{E}[Y_{0,i} \mid Y_{1}]\\
    &=\mathbb{E}[Y_{0,i} \mid Y_{1}]-\mathbb{E}[Y_{0,i} \mid Y_{1}]=0.
\end{split}
$}
\end{equation}
Therefore, Eq.~\eqref{eq:training_residual_5} can be simplified to the following factorization form under $\ell_2$ loss:
\begin{equation}
    \mathbb{E}[\mathcal{E}_{i}\mid Y_1]=w_{i}(Y_1)\cdot e_{i}(Y_1),
\end{equation}
which proves the Proposition~\ref{proposition:feactorization}.
\hfill $\blacksquare$

\subsection{Derivation of Training Residual}
By the law of total expectation and the tower property, the training residual:
\begin{equation}
\begin{split}
        \mathbb{E}[\mathcal{E}_{i}]&=\mathbb{E}[\mathbb{E}[\mathcal{E}_{i}\mid Y_1]\mid Y_{1,i}=0]P[Y_{1,i}=0]\\
        &+\mathbb{E}[\mathbb{E}[\mathcal{E}_{i}\mid Y_1]\mid Y_{1,i}\ne0]P[Y_{1,i}\ne 0].
\end{split}
\end{equation}
Using Proposition~\ref{proposition:feactorization}, this becomes:
\begin{equation}
\begin{split}
    \mathbb{E}[\mathcal{E}_{i}]&=\mathbb{E}[w_{i}(Y_{1})e_{i}(Y_{1})\mid Y_{1,i}=0]\cdot P[Y_{1,i}=0]\\
    &+\mathbb{E}[w_{i}(Y_{1})e_{i}(Y_{1})\mid Y_{1,i}\ne 0]\cdot P[Y_{1,i}\ne 0].
    \label{eq:unconditional_residual}
\end{split}
\end{equation}
Recall that $w_i(Y_1)$ is given in Proposition~\ref{proposition:wi(Y1)}:
\begin{equation}
\resizebox{\linewidth}{!}{$
        w_{i}(Y_{1})=(1-\frac{1-p_{i}}{1-p_{i}q_{i}}(1-\mathbb{E}[M_{Y,i}M_{1,i}\mid Y_{1}])) \mathbb{E}[M_{2,i}\mid Y_{1}, Y_{i}\ne 0],
$}
\label{eq:wi_appendix}
\end{equation}
We now derive $\mathbb{E}[\mathcal{E}_{i}]$ in Eq.~\eqref{eq:unconditional_residual} separately for disjoint and independent re-undersampling.

\textbf{\textit{Disjoint re-undersampling:}} 
If $Y_{1,i}\ne0$, then the location has already been assigned to the input subset and cannot appear in the supervision subset, so $\mathbb{E}[M_{2,i}\mid Y_{1}, Y_{i}\ne 0]=0$, and therefore $w_{i}(Y_{1})=0$.

If $Y_{1,i}=0$, we obtain $\mathbb{E}[M_{Y,i}M_{1,i}\mid Y_{1}]=0$, and this location must be sampled in the supervision subset if it is sampled in the initial undersampling, meaning that $\mathbb{E}[M_{2,i}\mid Y_{1}, Y_{i}\ne 0]=1$. Therefore, Eq.~\eqref{eq:wi_appendix} can be simplified to:
\begin{equation}
    w_{i}(Y_{1})=1-\frac{1-p_{i}}{1-p_{i}q_{i}}=\frac{p_i(1-q_{i})}{1-p_{i}q_{i}}.
\end{equation}

Bringing $w_i(Y_1)$ under these two cases back to Eq.~\eqref{eq:unconditional_residual} and considering $P[Y_{1,i}\ne0]=p_iq_i$, we have:
\begin{equation}
\begin{split}
        \mathbb{E}[\mathcal{E}_{i}]&=\mathbb{E}[\frac{p_i(1-q_{i})}{1-p_{i}q_{i}}e_{i}(Y_{1})\mid Y_{1,i}=0]\cdot (1-p_iq_i)+0\cdot p_iq_i\\
        &=p_i(1-q_{i})\cdot\mathbb{E}[e_{i}(Y_{1})\mid Y_{1,i}=0].
\end{split}
\end{equation}
This proves the Corollary~\ref{corollary:disjoint}.
\hfill $\blacksquare$

\textbf{\textit{Independent re-undersampling:}}
Under independet re-undersampling, $M_2$ is independent of $M_1$ given $Y$, we have:
\begin{equation}
    \mathbb{E}[M_{2,i}\mid Y_{1}, Y_{i}\ne 0]=P[Y_{2,i}\ne 0\mid Y_{i}\ne 0]=r_i.
\end{equation}

If $Y_{1,i}\ne0$, meaning that $\mathbb{E}[M_{Y,i}M_{1,i}\mid Y_{1}]=1$, Eq.~\eqref{eq:wi_appendix} simplifies to:
\begin{equation}
    w_{i}(Y_{1})=\mathbb{E}[M_{2,i}\mid Y_{1}, Y_{i}\ne 0]=r_i.
\end{equation}

If $Y_{1,i}=0$, we have $\mathbb{E}[M_{Y,i}M_{1,i}\mid Y_{1}]=0$, and Eq.~\eqref{eq:wi_appendix} becomes:
\begin{equation}
\begin{split}
        w_{i}(Y_{1})&=(1-\frac{1-p_{i}}{1-p_{i}q_{i}}) \mathbb{E}[M_{2,i}\mid Y_{1}, Y_{i}\ne 0]\\
        &=\frac{p_ir_i(1-q_{i})}{1-p_{i}q_{i}}.
\end{split}
\end{equation}

Similarly, substituting these expressions back into Eq.~\eqref{eq:unconditional_residual} gives:
\begin{equation}
\resizebox{\linewidth}{!}{$
\begin{split}
        \mathbb{E}[\mathcal{E}_{i}]&=\mathbb{E}[\frac{p_ir_i(1-q_{i})}{1-p_{i}q_{i}}e_{i}(Y_{1})\mid Y_{1,i}=0]\cdot (1-p_iq_i)\\
        &+\mathbb{E}[r_ie_{i}(Y_{1})\mid Y_{1,i}\ne 0]\cdot P[Y_{1,i}\ne 0]\\
        &=p_ir_i(1-q_i)\mathbb{E}[e_{i}(Y_{1})\mid Y_{1,i}=0]+p_iq_ir_i\mathbb{E}[e_{i}(Y_{1})\mid Y_{1,i}\ne0].
\end{split}
$}
\end{equation}
This proves the Corollary~\ref{corollary:independent}.
\hfill $\blacksquare$

\subsection{Variance Interpretation of Cross-consistency}
\label{sec:proof_variance}
To interpret the effect of cross-consistency, we examine the prediction bias defined in Eq.~\eqref{eq:def_ei} under the two re-undersampled inputs. For a fixed k-space location $i$, we have:
\begin{equation}
    \begin{split}
        e_{i}(Y_1):=f_i(Y_1)-f^{\star}_i(Y_1),\\
        e_{i}(Y_2):=f_i(Y_2)-f^{\star}_i(Y_2).
    \end{split}
\end{equation}
where $f^{\star}_i(\cdot)$ denotes a Bayes-optimal predictor under the chosen loss.

We now consider the joint supervision via the cross-consistency loss, and define the averaged prediction bias as:
\begin{equation}
    \bar{e_{i}}=\frac{1}{2}(e_{i}(Y_1)+e_{i}(Y_2)).
\end{equation}
Its variance is obtained by direct expansion:
\begin{equation}
\resizebox{\linewidth}{!}{$
\begin{split}
    \mathrm{Var}(\bar{e_{i}})&=\mathrm{Var}(\frac{1}{2}(e_{i}(Y_1)+e_{i}(Y_2)))\\
    &=\frac{1}{4}\mathrm{Var}(e_{i}(Y_1))+\frac{1}{4}\mathrm{Var}(e_{i}(Y_2))+\frac{1}{2}\mathrm{Cov}(e_{i}(Y_1),e_{i}(Y_2)).
\end{split}
$}
\label{eq:variace_average_error}
\end{equation}

Eq.~\eqref{eq:variace_average_error} shows that the variance of the averaged prediction bias depends on the pathwise variances and their covariance. Therefore, the potential gain from cross-consistency depends on the correlation between the two pathwise errors.

Since the two inputs are re-undersampled from the same acquired k-space and processed by the same reconstruction network, the variances of the two prediction biases are expected to be comparable in practice. Under this heuristic approximation:
\begin{equation}
    \mathrm{Var}(e_{i}(Y_1))\approx \mathrm{Var}(e_{i}(Y_2))\approx\sigma_i ^{2},
\end{equation}
Eq.~\eqref{eq:variace_average_error} becomes:
\begin{equation}
    \mathrm{Var}(\bar{e_{i}})\approx\frac{1}{2}\sigma_i ^{2}+\frac{1}{2}\mathrm{Cov}(e_{i}(Y_1),e_{i}(Y_2)),
\end{equation}
and the Cauchy-Schwarz inequality gives:
\begin{equation}
    \left | \mathrm{Cov}(e_{i}(Y_1),e_{i}(Y_2)) \right | \le \sqrt{\mathrm{Var}(e_{i}(Y_1))\mathrm{Var}(e_{i}(Y_2))}\approx \sigma_i ^{2}. 
\end{equation}
Therefore,
\begin{equation}
    \mathrm{Var}(\bar e_i)\lesssim \sigma_i^2.
    \label{eq:variance_reduction}
\end{equation}

Eq.~\ref{eq:variance_reduction} suggests a potential stabilizing benefit of cross-consistency loss. When the errors of two pathways are weakly correlated, averaging can reduce variance. When they are strongly positively correlated, the benefit decreases. Importantly, this benefit arises without any architectural change, only from leveraging both available supervision information.

\section*{Acknowledgments}
The work was supported by the Deutsche Forschungsgemeinschaft (DFG, German Research Foundation) under Germany’s Excellence Strategy – EXC 2064/1 – Project number 390727645.
This work was supported by the de.NBI Cloud within the German Network for Bioinformatics Infrastructure (de.NBI) and ELIXIR-DE (Forschungszentrum Jülich and W-de.NBI-001, W-de.NBI-004, W-de.NBI-008, W-de.NBI-010, W-de.NBI-013, W-de.NBI-014, W-de.NBI-016, W-de.NBI-022).

\bibliographystyle{IEEEtran}
\bibliography{refs}

\clearpage
\onecolumn
\section*{Supplemental Figures}
\setcounter{equation}{0}
\setcounter{figure}{0}
\setcounter{table}{0}
\setcounter{page}{1}
\makeatletter
\renewcommand{\theequation}{S\arabic{equation}}
\renewcommand{\thefigure}{S\arabic{figure}}
This supplementary document provides additional figures supporting the main paper.

\begin{figure*}[h]
\centering
\includegraphics[width=\linewidth]{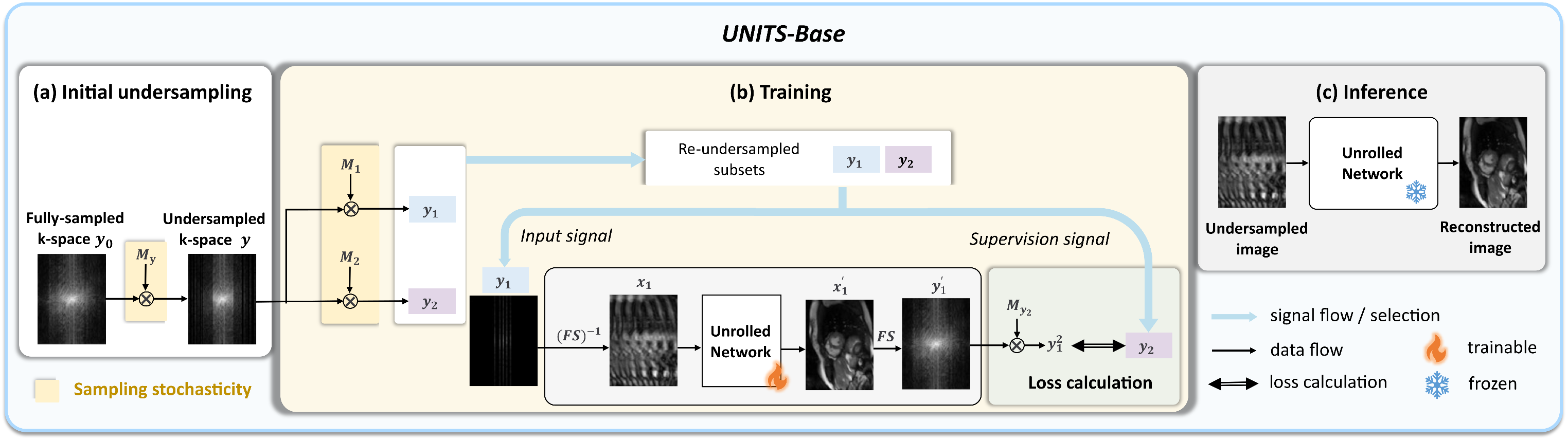}
\caption{\textbf{Visualization of \textit{UNITS-Base} within the UNITS framework.} (a) Initial undersampling: at each training step, a random mask $M_{y}$ with variable acceleration rates and random seeds is generated to acquire undersampled data $y$. (b) Training: the acquired k-space $y$ is re-undersampled into two independent subsets $y_{1}$ and $y_{2}$ with randomized ratios. The subset $y_{1}$ serves as input, while $y_{2}$ provides supervision through loss calculation. The network is a physics-based unrolled network. (c) Inference: the trained unrolled network directly reconstructs undersampled images.}
\label{fig:units_base}
\end{figure*}

\begin{figure*}[h]
\centering
\includegraphics[width=\linewidth]{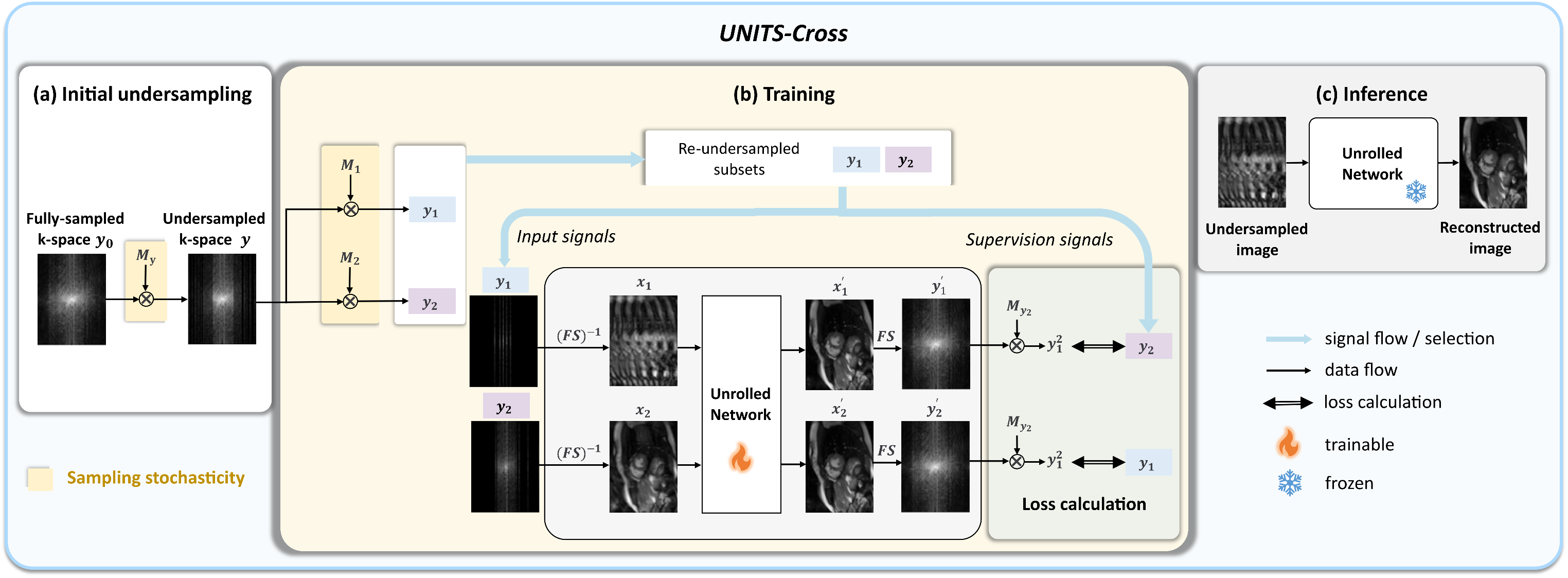}
\caption{\textbf{Visualization of \textit{UNITS-Cross} within the UNITS framework.} (a) Initial undersampling: at each training step, a random mask $M_{y}$ with variable acceleration rates and random seeds is generated to acquire undersampled data $y$. (b) Training: the acquired k-space $y$ is re-undersampled into two independent subsets $y_{1}$ and $y_{2}$ with randomized ratios. Both subsets are used as inputs and supervision signals: the unrolled network reconstructs $y_{1}$ and $y_{2}$ in parallel, and each subset provides supervision for the other during loss calculation. (c) Inference: the trained unrolled network directly reconstructs undersampled images.}
\end{figure*}

\begin{figure*}[h]
\centering
\includegraphics[width=\linewidth]{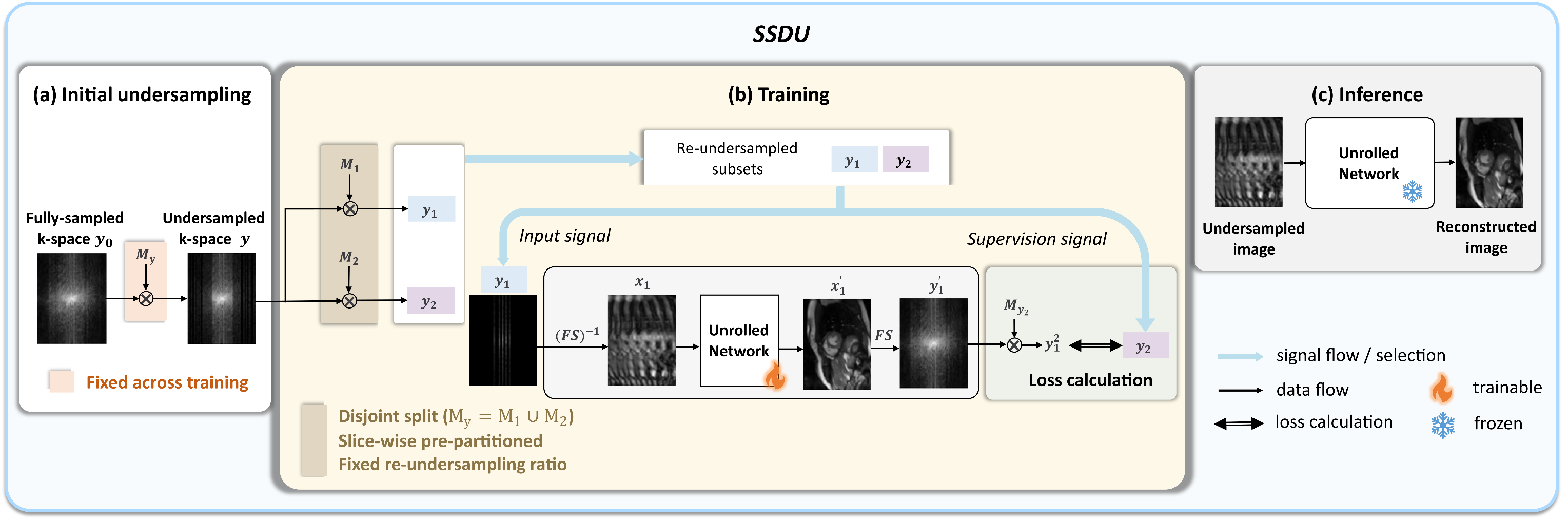}
\caption{\textbf{Visualization of \textit{SSDU}~\cite{yaman2020self} within the UNITS framework.} (a) Initial undersampling: a fixed mask $M_{y}$, unchanged during training, is applied to acquire undersampled data $y$. (b) Training: for each slice, the acquired k-space is pre-partitioned into two disjoint subsets $y_{1}$ and $y_{2}$ with a fixed ratio. The subset $y_{1}$ serves as input, while $y_{2}$ provides supervision through loss calculation. The network is a physics-based unrolled network. (c) Inference: the trained unrolled network directly reconstructs undersampled images.}
\end{figure*}

\begin{figure*}[h]
\centering
\includegraphics[width=\linewidth]{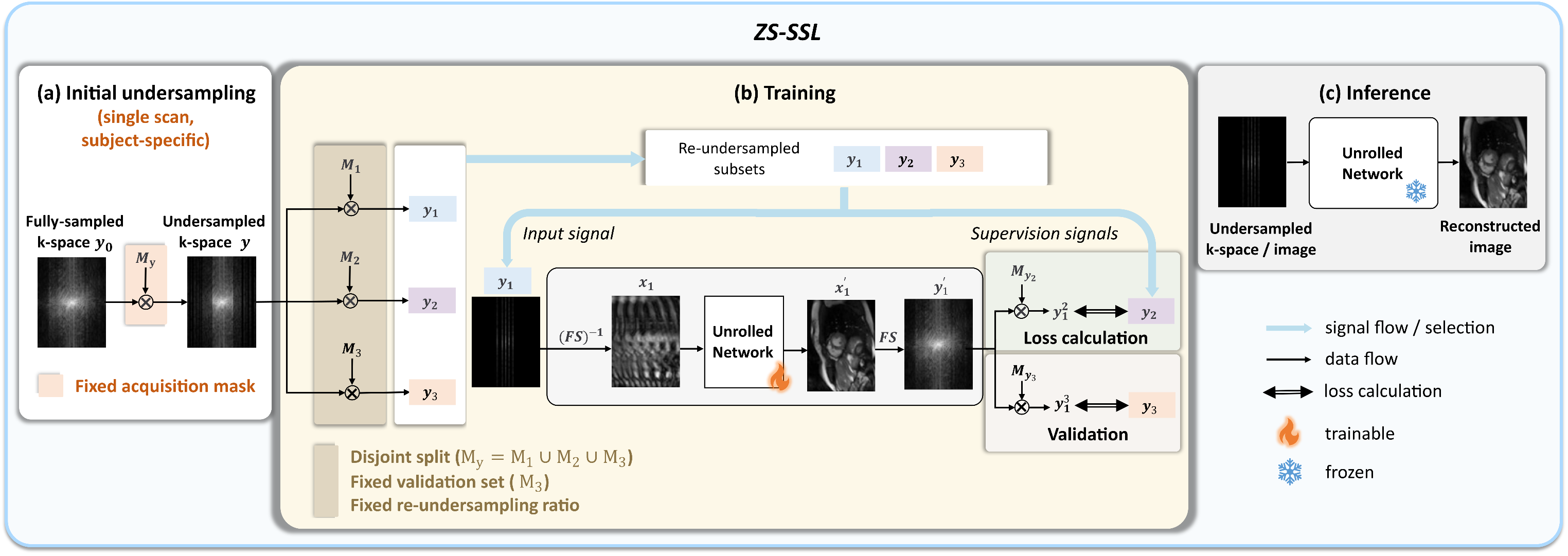}
\caption{\textbf{Visualization of \textit{ZS-SSL}~\cite{yaman2021zero} within the UNITS framework.} (a) Initial undersampling: a single subject-specific undersampled scan $y$ is acquired with a fixed acquisition mask. (b) Training: the acquired k-space is re-undersampled into three disjoint subsets $y_{1}$, $y_{2}$, and $y_{3}$ with a fixed ratio. The subset $y_{1}$ serves as input, $y_{2}$ provides supervision through loss calculation, and $y_{3}$ is held out as a fixed validation set for automated early stopping. (c) Inference: the trained unrolled network directly reconstructs undersampled images of the same scan.}
\end{figure*}

\begin{figure*}[h]
\centering
\includegraphics[width=\linewidth]{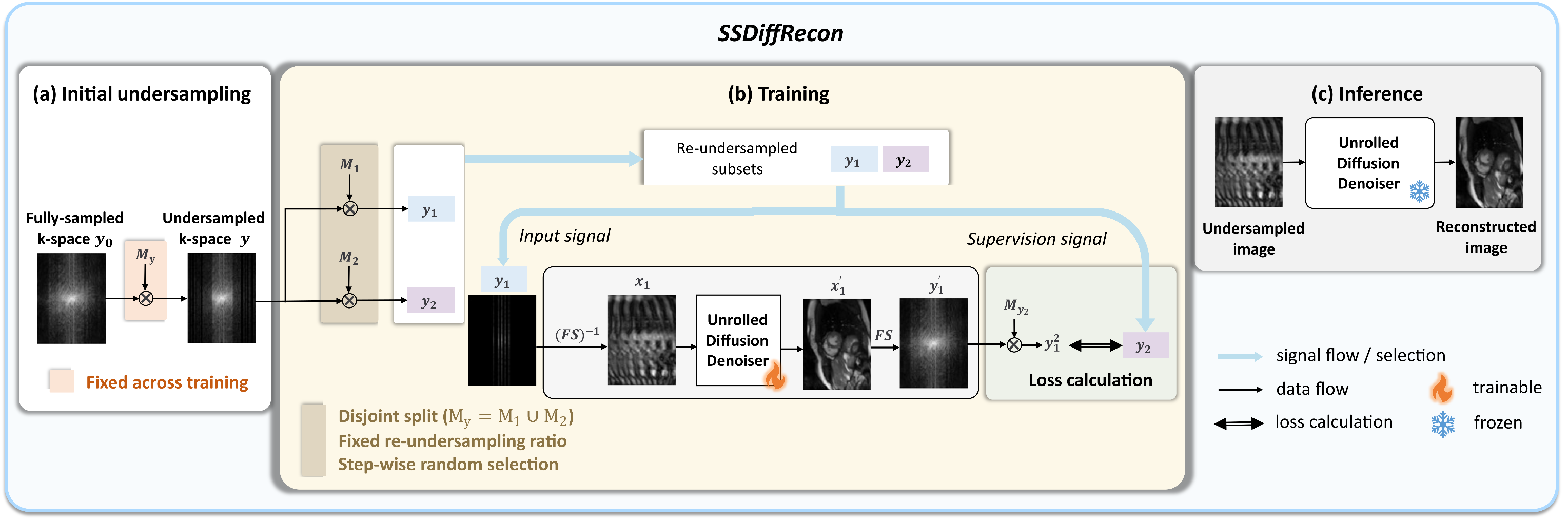}
\caption{\textbf{Visualization of \textit{SSDiffRecon}~\cite{korkmaz2023self} within the UNITS framework.} (a) Initial undersampling: a fixed mask $M_{y}$, unchanged during training, is applied to acquire undersampled data $y$. (b) Training: the acquired k-space is re-undersampled into two disjoint subsets randomly at each training step with a fixed ratio. The subset $y_{1}$ serves as input to the unrolled diffusion denoiser, while $y_{2}$ provides supervision through loss calculation. (c) Inference: the trained denoiser directly reconstructs undersampled images.}
\end{figure*}

\end{document}